\documentclass[twocolumn]{mn2e}
\usepackage{lipsum}
\usepackage{etoolbox}
\usepackage{epsfig}
\usepackage{epstopdf}
\usepackage{subfloat}
\usepackage{textcomp}
\usepackage{subfigure}
\usepackage{multirow}
\usepackage{graphicx,times}  
\usepackage{soul}
\usepackage{natbib}
\usepackage{amssymb,amsmath}
\usepackage{xspace}
\usepackage{comment}

\newcommand{\vt}{{\vec \theta}}
\newcommand{\vU}{{\vec{ U}}}

\newcommand{\Cl}{C_{l_g} }

\def\HI{{H~{\sc i} }}

\begin{document}

\title[Strong lensing to probe post reionization \HI]{Combined lensed estimator to probe the post reionization \HI power spectrum}
\author[Urvashi Arora and Prasun Dutta ]
{Urvashi Arora$^{1}$\thanks{Email: urvashi.rs.phy17@itbhu.ac.in}, 
Prasun Dutta$^{1}$\thanks{Email:pdutta.phy@itbhu.ac.in},  
\\$^{1}$ Department of Physics, IIT (BHU) Varanasi, 221005  India. 
}
\maketitle
\begin{abstract}
In the post-reionization era, the baryons assembled into the protogalaxies and eventually the present population of the galaxies evolved through merger and evolution. In this work, we discuss a possible probe of the statistical distribution and evolution of the \HI density in the post reionization era. We introduce an estimator of the \HI power spectrum from the post reionization universe by observing it through the strong gravitational lenses by the nearby galaxy cluster. We also analytically calculate the uncertainties associated with the estimates of the post-EoR power spectrum for the discussed estimator. We access the efficacy of this estimator in the context of $19$ galaxy clusters for which the lensing potential has been estimated earlier by various authors. We find that by combining the lensed power spectrum through  eight of these  cluster lenses, it is possible to estimate the post-reionization \HI power spectrum  at five-sigma significance for angular multipoles $<4000$ for a uGMRT observation of $16$ MHz bandwidth from redshifts of $1.25, 1.5$ with a total of $400$ hours of observation. With the same setup, for a redshift of $3.0$, we need  $200$ hours of total observation time. The estimator also suppresses the diffused galactic foreground, though, the latter is still a dominant contributor to the overall signal and hence need to be estimated and mitigated. We discuss the merits and demerits of the estimator.
\end{abstract}

\section{Introduction}


In the present understanding of cosmological structure formation, the primordial over-densities in dark matter grew through gravitational clustering \citep{1980lssu.book.....P,1996IAUS..173...55P}. The over-densities eventually became non-linear, got isolated from the cosmological expansion and collapsed further \citep{1974ApJ...187..425P,10.1093/mnras/183.3.341}. In this era, Baryons, mostly atomic hydrogen ( henceforth \HI) and Helium followed the dark matter evolution   \citep{2001PhR...349..125B,2005MNRAS.356.1519B,2006PhR...433..181F,2013MNRAS.429.2537M}. To the best of our knowledge, the assembly of baryons gave rise to the first luminous objects around the redshifts of $30$ \citep{1997ApJ...474....1T,2001PhR...349..125B}. Radiation from these objects started to ionised the universe and the universe entered into the reionization era \citep{2001PhR...349..125B,2013ASSL..396...45Z}. Observations of Lyman $\alpha$ spectra suggests that by the redshift of $6$ the universe was completely ionised \citep{2015MNRAS.454L..76M,2019MNRAS.485L..24K,2020MNRAS.494..703W}. 
At the later redshifts, the major structural change in the universe happens through the formation of galaxies, their mergers and evolution \citep{2010gfe..book.....M}. 

In the post-reionization era \HI is mostly trapped in the galaxies, like Damped Lyman-$\alpha$ absorbers, where they are shielded from the ionizing radiation \citep{1965ApJ...142.1633G,2000eaa..bookE2141L, 2006PhR...433..181F, 2012RPPh...75h6901P}. However, there is a significant amount of gas left  in the IGM \citep{2016ARA&A..54..313M}. This gas, as  time evolves, falls into the galaxies \citep{1993MNRAS.264..201K,1998ApJ...498..504B}. Star formation and its feedback also start to play important role in the evolution of the large scale \HI density \citep{2009MNRAS.400..154B,2018PASP..130i4101Z}. Though there is some consensus on how the global \HI densities evolved in the post reionization universe \citep{2019ApJ...882L...7B,2020Natur.586..369C}, much to be explored to understand its distribution through an observational probe\citep{2010MNRAS.407..567B,2020MNRAS.498.3275A}. Recently, \citet{2019MNRAS.485L..24K,2020MNRAS.494..703W}  have shown that the spatial fluctuations in the Ly$\alpha$  at the redshifts of around 5-6 can be a result of late and patchy reionization that leaves isolated neutral islands near the end of reionization.  This is also expected to change the otherwise accepted view of the \HI statistics in the post reionization universe.

Observation of the individual \HI clumps from the post reionization universe is beyond the reach of the present telescopes. The distribution of the \HI and its statistical properties can be probed by the integrated radio emission from unresolved gas clouds, this technique is known as intensity mapping \citep{1997ApJ...475..429M,2001JApA...22...21B,2014JCAP...09..050V}. Radio interferometers are the instrument of choice for intensity mapping experiments. \cite{2005MNRAS.356.1519B} showed that directly observed quantity visibility from the radio interferometers can be used  to probe statistics of the 21 cm signal from various redshifts. \cite{2010MNRAS.407..567B} used N-body simulation to put constraints on the neutral hydrogen contents in the dark matter halos and gives the possibility of observation of the statistical distribution of \HI at an angular  scale range of $1.5'$ to $6.5'$ with the Giant Meterwave Radio Telescope (henceforth GMRT \footnote{GMRT: Giant Meterwave Radio Telescope, National Centre for Radio Astrophysics, TIFR, India \citep{swarup...1991}}) at  a redshift  of $z \sim 1.3$ with an observational time of $400$ hours and bandwidth of $16$ MHz.  \cite{2014JApA...35..157A} estimated that observation of  \HI power spectrum at a redshift of $z= 3.35$ is possible for an angular scale of $11'$ to $3^{\circ}$ with a three-sigma significance for an observation time of $1000$ hours and bandwidth of $30$ MHz with the ORT (henceforth ORT\footnote{ORT: Ooty Radio Telescope, National Centre for Radio Astrophysics, TIFR, India \citep{1971NPhS..230..185S}}), Phase II. \cite{2011MNRAS.410.1130G,2017JCAP...04..001C} used the cross-correlation between 21-cm intensity mapping and Ly$\alpha$ forest in order to constraints the \HI content in the galaxies in the post reionization universe.

A major problem in observing the redshifted \HI signal is the presence of diffuse galactic synchrotron emission from our galaxy at the observing frequencies, the so-called foreground emission \citep{2008MNRAS.389.1319J,2009ApJ...695..183B}.  Several methods for foreground suppression and removal \citep{2011MNRAS.418.2584G, 2012ApJ...749..164C, 2016MNRAS.463.4093C}
and foreground avoidance \citep{2016ApJ...818..139T,2018MNRAS.479.2767Y} has been discussed. Observationally, there have been several attempts to quantify the properties of the foreground signal by \citet{2009MNRAS.399..181P,2011MNRAS.418.2584G, 2019MNRAS.487.4102C}. \cite{2016ApJS..222....3Z} use the Bayesian-based semi-blind component separation approach to remove foreground contamination from interferometric observation. \cite{2020MNRAS.495.2813G} uses the Gaussian Process Regression to model both foreground emission and instrumental systematics in $\sim2$ hours observation from the HERA (henceforth HERA \footnote{HERA: Hydrogen Epoch of Reionization Array, South Africa \citep{2017PASP..129d5001D}}). 

 \cite{2015MNRAS.448.2368P, 2018MNRAS.474.1787R} demonstrate the possible use of weak gravitational lensing for 21-cm intensity mapping using SKA-mid (henceforth SKA \footnote{SKA: Square Kilometer Array, Australia and South Africa \citep{8105425, 8105424} }) and SKA-low radio telescope  in  order  to  observe  the  lensed  cosmological  21-cm  signal. \cite{2001ApJ...557..421S} discuss the possibility of detecting HI signal from high redshift gas clouds by strong gravitational lensing from the cluster lenses using the GMRT. This is further explored in \cite{2015MNRAS.452L..49D} for present and future radio telescopes. \cite{2019MNRAS.484.3681B} reports the first observation of  lensed \HI from a galaxy at redshift $0.4$ using the  GMRT. \citet{2018MNRAS.476.4383D} has detected strongly lensed highly star-forming galaxies using ALMA (henceforth ALMA \footnote{ALMA: Atacama Large Millimeter/Submillimeter Array, Chile \citep{2009IEEEP..97.1463W}}) observations.
 
There have been several approaches and studies to derive the lensing potential of galaxy clusters using optical observations \citep{2010MNRAS.404..325R, 2014MNRAS.444..268R, 2018ApJ...859..159C, 2020ApJS..247...12S}. \cite{2005MNRAS.359..417S, 2007NJPh....9..447J, 2010MNRAS.404..325R, 2015MNRAS.452.1437J, 2018ApJ...859..159C} uses different approaches to model the lensing potential using multiple images of the background galaxies. These lens models are then used to reconstruct the morphological  properties of the  galaxies at higher redshifts and infer their star-formation and dynamical properties \citep{2018MNRAS.481.1427S, Chiriv__2020}. These studies have enriched our knowledge of the lensing potential over the last few decades.

Recently, \citet{2020MNRAS.498.3275A} explored the idea of using strong gravitational lensing of statistical distribution of \HI to probe the redshifted 21-cm signal from the post reionization universe. They find that the strong lensing by individual galaxy cluster enhances the \HI power spectrum at the scale of the cluster or lower. Since the lensing only magnify the background signal, not the foreground, the effect of the foreground is less prominent in the lensed  \HI signal. They also discuss that owing to the structure of the lensing potential the effect of  lensing is effective only in a limited region of the sky. This requires a more detailed investigation 
to check the feasibility of this method to determine the redshifted 21-cm signal from the post reionization era.

In this work, we construct an unbiased power spectrum estimator with the lensed visibilities and explore the detection criteria of redshifted 21-cm signal from the post reionization era using strong gravitational lensing using a single lens as well as by the combination of signals from many lenses. The estimator is designed following the gridded estimator by  \citet{2014MNRAS.445.4351C}. Here we assume that the lensing potential is well known and explore the uncertainties in the measurement of the power spectrum only arising from sample variance and instrumental errors. Section~2 introduces the lensed power spectrum estimator and its uncertainties, we discuss one possible implementation of this estimator in secition~3. In section~5 we present an estimation of signal to noise for observation with parametric potentials of a few lenses from the literature. We discuss the results and conclude in section~6.

\section{Combined Lensed Estimator}
\subsection{21cm power spectrum}
Let  $\delta I(\vt, z_s )$ be the specific intensity fluctuation of \HI emission  originated at the redshift $z_s$  from a direction $\vt$ in the sky with respect to the centre of the  field of view of observation.  We define $\Delta \tilde I(\vU)$ as the Fourier transform of $\delta \tilde I(\vt, z_s)$ 
\begin{equation}
 \Delta \tilde{I}(\vU, z_s)= \int d\vt \, \delta I(\vt, z_s) e^{-i2\pi \vU.\vt}.
 \label{eq:PS}
\end{equation}
The vector $\vU$, we refer as `baseline', is the Fourier conjugate to the vector $\vt$. Assuming the specific intensity fluctuations to be statistically isotropic, angular power spectrum $C_l$   of the \HI 21-cm brightness fluctuations at a multipole $l = 2 \pi U $ , ($U = \mid \vU \mid$) can be written as 
\begin{equation}
\langle \Delta \tilde{I}(\vU, z_{s})\Delta \tilde{I}^*(\vU, z_{s}) \rangle =  \delta_{2D}(\vU - \vU') \left(\frac{\partial B}{\partial T}\right)^2 C_l.
\label{eq:CORR}
\end{equation}
Here we use angular brackets $< >$ to represent the ensemble average  and $\delta_{2D}$ is the two dimensional Dirac delta function. The angular power spectrum $C_l$ probes  the matter power spectrum at the source redshift $z_{s}$ (see for example \citet{2014MNRAS.445.4351C}).

\subsection{Effect of Strong Lensing}
\cite{2020MNRAS.498.3275A} has explored the modification of the 21-cm power spectrum from a strongly lensed  region. We can write the brightness temperature fluctuation of 21-cm emission originating from redshift $z_s$ through a strong lens at redshift $z_L$ as
\begin{equation}
\delta I_{L}(\vec{\theta}, z_{s}, z_{L})= \int d \vec{\theta'}   \delta I(\vec{\theta'}, z_{s}) \, G_{L}(\vec{\theta}-\vec{\theta'}, z_{s}, z_{L} ),    
\end{equation}
where $ G_{L}(\vec{\theta}-\vec{\theta'}, z_{s}, z_{L} )$ is the point spread function of the gravitational lens. For a gravitationally lensed signal, radio interferometers measure  the visibility function $V(\vU, z_s, z_L)$  as
\begin{equation}
V(\vec{U}, z_s, z_L) = S_L(\vec{U}, z_s, z_L) V_s(\vec{U}, z_s) +N(\vec{U}).   
\label{eq:Vsdef}
\end{equation}
Here $S_L$ is the lens sampling function and is the Fourier transform of $G_{L}$ in the same sense as in eqn~(\ref{eq:PS}).  The quantity $N(\vU)$  denotes the complex system noise at each baseline. The quantity $V_s(\vec{U}, z_s)$ is related to $\Delta\tilde I(\vec{U}, z_s)$ as
\begin{equation}
V_s(\vec{U}) = \int d\vec{U} \, \tilde{a}(\vec{U}-\vec{U}') \Delta\tilde I(\vec{U})
\label{eq: VISI}
\end{equation}
where $\tilde{a}(\vec{U})$ is the antenna beam pattern. We have not explicitly written the redshift dependence henceforth. This quantity $V_s$ represents the visibility in absence of any lens and is measured by an interferometer with no noise. In practice, the antenna beam pattern can be complex. However, for most of the interferometers, a Gaussian function approximates the antenna beam \citep{2014MNRAS.445.4351C}. In this work, we assume  the antenna beam pattern to have the following form and use it henceforth.
\begin{equation}
\tilde{a}(\vec{U}) = \pi \theta_0^{2} \exp \left[ - \pi^2 \theta_0^2 U^2 \right ],
\end{equation}
where $\theta_0$ defines the field of view of observation. We define 
\begin{equation}
V_{2S}(\vU, \Delta \vU) = \langle V_S(\vU) V_S^*(\vU+\Delta \vU) \rangle. 
\end{equation}
Using  eqn~(\ref{eq: VISI}) and eqn~(\ref{eq:CORR}) we can write above   as
\begin{equation}
V_{2S}(\vU, \Delta \vU) = \left(\frac{\partial B}{\partial T}\right)^2 \, \int d\vU \, \tilde{a}(\vU-\vU') \, \tilde{a}^*(\vU+\Delta \vU -\vU') \, C_l. 
\end{equation}
In the limit $\mid \Delta \vU \mid << 1/\theta_0$, we may write above as
\begin{equation}
V_{2S}(\vU, \Delta \vU) = V_{20}\, D(|\Delta \vU|) \, C_l   
\label{eq:defV2S}
\end{equation}
where $D(\Delta \vU) = \exp \left[ - \pi^2 \theta_0^2 \mid \vU - \vU' \mid ^2  \right ] $ and $V_{20} = \frac{\pi \theta_0^2}{2}\left(\frac{\partial B}{\partial T}\right)^2  $. 

\subsection{Power Spectrum Estimator}
Interferometers sample visibilities at discrete baseline positions given by instantaneous projected separation of antenna pairs in the sky plane in units of observed wavelengths. We denote one sample of the  visibility function measured at a baseline $\vU$ through a particular  gravitational lens as $V_i$. Using eqn.~\ref{eq:Vsdef} we can write $V_i$ in terms of the lens sampling function $S_{Li}$ at that baseline for the same lens, $V_{Si}$, and system noise $N_i$ as
\begin{equation}
V_i = S_{Li} V_{Si}+ N_i.
\label{eq:LS}
\end{equation}
\citet{2020MNRAS.498.3275A} have shown the effect of lensing by a single cluster and demonstrated the increase in the observed signal in presence of lensing.  However, they do not discuss an unbiased estimator of the 21-cm power spectrum. Moreover, one  limitation in their method is that strong lensing enhances the signal only near the caustics making the enhanced signal  sample variance limited. They show that  the lensing sampling function enhances the visibility correlation only when its modulus is above unity and such baseline positions for a single gravitational lens can be limited.  Since the brightness temperatures are assumed to be homogeneous and isotropic, observation of the specific intensity fluctuation at the different direction in the sky at the same redshift $z_s$  can be used to probe the statistical nature of the signal $V_S$ at that redshift.  Hence, we may combine estimates of the power spectrum by several galaxy clusters to effectively enhance the sampling of the baseline plane. In this approach, we invert the effect of lensing to estimate the power spectrum of sky brightness distribution unbiasedly. Here we assume that the lensing sampling function is known from an earlier study to considerable accuracy.

We construct the angular power spectrum estimator in the following way\footnote{This is a modified version of the estimator discussed in give \cite{2014MNRAS.445.4351C}}. 
We first grid the baseline plane with grid size $\sim 1/\theta_0$. We collect visibilities from one lens in the sampled baseline grids.   The mean of the angular power spectrum and its uncertainties are then estimated in each grid. We define, for each grid, 
\begin{equation}
E_L(g) = \frac{1}{V_{20}  P} \sum _{i,j} w_{ij} V_i V_j^*
\label{eq:est}
\end{equation}
where the summations are over all the measured visibilities in a grid. Here $D_{ij}$ is the function $D(\Delta U)$ defined in eq~\ref{eq:defV2S} with $\Delta U$ being the magnitude of the separation of $i^{th}$ and $j^{th}$ baselines. The weight factor $w_{ij}$ has the form  $w_{ij} = \Delta_{ij} k_{ij}$, where $\Delta_{ij} = (1- \delta_{ij})$, $\delta_{ij}$ is the Kronecker delta and $k_{ij}$ is chosen to enhance the signal to noise of the estimator. In this work, we choose $k_{ij} = (S_{Li}S^{*}_{Lj})^{-1}$ and  $P$ is $Trace[P_{ij}]$ when $P_{ij} = \sum_{k}\Delta_{ik} D_{kj}$.  

It is to be noted that, in a typical interferometer, the noise in each baseline is expected to be higher than the sky signal at that baseline. Hence, if we consider visibility correlation between the same baseline, the noise term always dominates and gives rise to a bias in the estimator. The term $\Delta_{ij}$ ensures that the same baseline correlations are not taken in $E_L(g)$. Noise in a typical interferometer is also not correlated across the baselines. Hence, for an interferometer, with per visibility mean square system noise as $\sigma^2_N$, $\langle N_i N_j^* \rangle = 2 \sigma^2_N \delta_{ij}$.  In presence of enough visibility measurements in the grid $g$ and absence of foreground, the  quantities $E_L(g)$ gives an unbiased estimate of the angular power spectrum $\Cl$ in the grid $g$  defined by $\vU_{g} = (u_g, v_g)$ as $u_g = \sum _{i} P_{ii} u_{i} /P , \ \ v_g = \sum_{i} P_{ii} v_{i}/P$.

 The baseline grid with $ \vU_g = (u_g, v_g)$ correspond to the multipole $l_g = 2\,  \pi\, \sqrt{u_g^2 + v_g^2} $. There is only one estimate of the angular power spectrum in a given grid. Uncertainty in this estimate, in the same grid is given as
 \begin{equation}
 \label{eq:var}
 \sigma^2_{g}  =  \frac{\left[ \Cl^2 \sum \limits_{i,j} P_{ij} P_{ji}  + \eta \Cl \sum \limits_{i, j} P_{ij} \Delta_{ji} k_{ii}  + \eta^2  \sum \limits_{i,j} k_{ii} \Delta_{ij} k_{jj} \right]}{P^2},
 \end{equation}
 where $\eta = \frac{2\sigma_N^2}{V_{20}}$.
 Note that, with the choice of $k_{ij}$ here, we have made the estimates of $C_l$ in the grids  independent of the lensing sampling function. However, the uncertainties in the estimates in each grid depend on the grid itself.  
 
 If observations of the lensed redshifted 21-cm signal are done for directions of multiple cluster lenses, then we collect all estimates of $\Cl$ and $\sigma^2_{g} $ in the same grids in the baseline plane. For the grids where with contribution from more than one cluster lenses we average over the estimates of $\Cl$ and quadrature average over the estimates of $\sigma^2_{g}  $. The azimuthally averaged estimates of the  angular power spectra $C_l$ and its variance $\sigma^2_l$ in a bin representing angular multipole $l$ can be written as
 \begin{equation}
 C_l = \sum_g \Cl /N_G, \ \ \ \ \ \ \ {\rm and} \ \ \ \  \sigma^2_l = \sum_g \sigma^2_{C_g}/N_G^2,
 \label{eq:azcl} 
 \end{equation}
 where the summation is over the number of grids $N_G$ in the annulus with estimates of $\Cl$ and the above measurements are at $l = \sum_g l_g /N_G$.

\section{Implementation of the Estimator}
The basic idea of the combined lens estimator and its variance is discussed in section~2. Here we discuss the implementation of the estimator through the following steps.

\subsection{Calculation of $S_L$ in the observed baseline positions}
In this work, we assume that we are using the gravitational lenses with sufficiently well-constrained lensing potential. Estimation of the lensing sampling function $S_L$ from the projected potential of the lenses is discussed in \cite{2020MNRAS.498.3275A}. We estimate the magnification function of the lenses from their known projected potential. For the cluster lenses, the extent of the  lensing potentials is significantly large as compared to the observing wavelengths. Hence the magnification function of the lens can be approximated as the point spread function $G_L$ \citep{10.1093/ptep/pty119}. We first estimate $G_L$ in  a grid in the sky plane. We choose the grid size based on the angular resolution of the telescope. Extend of the grid in the sky plane, over which $G_L$ is calculated, is chosen  such that the lens with the largest extent has unit magnification at the edge of the grid and beyond. While observing, an interferometer samples the visibility function at particular baseline positions given by the instantaneous projected separation of antenna pairs in the sky plane in terms of the wavelength of observation. The sampling of the visibility function in the baseline plane by a typical interferometer depends on its location,  antenna configuration, declination of the source and observing time. For all the lenses we estimate the lensing sampling function in the observed baseline  positions by calculating non-uniform Fast Fourier Transform of $G_L$ of the particular lens.  If for a given lens, a particular baseline has  $S_L<1$, lensing enhances the noise in the angular power spectrum estimate.  We choose a threshold value $S_T$ for the modulus of lensing sampling function.  We use measurements from a particular pointing in a grid only if  $\mid S_L \mid > S_{T}$ in that grid for the pointing. We always choose $S_T \ge 1$, for every grid, to reduce noise enhancement by lensing. Note that this further changes the sampling of the baseline plane and hence the term ``Lensing sampling function" is used for $S_L(\vU)$.

\subsection{Gridding the observed lens visibilities} 
We construct a grid in the baseline plane. Since every interferometer can measure to a maximum baseline, we need to only grid the baseline plane to the maximum baseline $U_{M}$.  Note that, in presence of the term $D(\mid \Delta \vU \mid)$ in eqn~\ref{eq:defV2S}, the visibility correlation $V_{2S}$ drops rapidly as $ \mid \Delta \vU \mid $ increases beyond $1/\theta_0$. If the caustics of the lens has an extension smaller than the field of view of the telescope, the effective value of $\theta_0$ may be taken as the extent of the caustics itself. We choose the grid-size  of the baseline grids as $U_m = 1/(\pi \, \theta_0)$. This ensures that the visibility correlations contribute significantly to the measurements of the angular power spectrum in each grid.  We label the visibility measurements  based on the baseline grid and the  lens. For a given lens, if the number of measurements of visibilities in a given grid  $N_b$ is rather less, an estimate of the angular power spectrum in that grid is not statistically significant. In fact to estimate the power spectrum, we need at least two baselines in a given grid. We define  a threshold number  $N_T>2$, such that we use a grid, only if $N_b > N_T$.  The grids which have $N_b < N_T$ are discarded.  We use the measurements in the rest of the grids for further analysis. Note that increasing both $N_T$ and $S_T$ decreases uncertainties in each grid. However, it also decreases the number of grids $N_G$ in a given annulus. Hence, an optimum choice is to be made while deciding on $S_T$ and $N_T$. 

\subsection{Estimates of  $C_{l}$ and its variance} 
 Given a lens, we  estimate the angular power spectrum in the grids where $N_b > N_T$ by directly using the estimator as given in eqn~\ref{eq:est}.  We also calculate the value of $l_g$ corresponding to the grid. An estimate of the variance in each of these grids are obtained using eqn~\ref{eq:var}.  Estimation of the quantity $\sigma_N$ will be discussed in the next section.
 
We collect contributions from all lenses for $l_g,  \Cl$ and $\sigma^2_g$ in the baseline plane grids. 
We divide the baseline plane in azimuthal bins with their boundaries in logarithmic intervals. For each bin,  we estimate the azimuthally average angular power spectrum and its uncertainty following definition in eqn~\ref{eq:azcl}. This gives us a measurement of the 21-cm angular power spectra and its uncertainty as a function of the angular multipole.

\section{Simulations and Results}
In this section, we demonstrate the effectiveness of the angular power spectrum estimator defined above by using a fiducial model for the 21-cm angular power spectrum and parametric models of known lenses from optical studies.
\subsection{21 cm Angular Power spectrum} 
Models of the post reionization  21-cm angular power spectra can be found in \citet{2005MNRAS.356.1519B}, \citet{2014MNRAS.445.4351C} \citet{2016MNRAS.460.4310S} and \citet{2018MNRAS.476...96S}. In this work, we use the semi-analytical model of the 21-cm angular power spectrum adopted from the work by \citet{2016MNRAS.460.4310S} and \citet{2018MNRAS.476...96S}. 
\begin{eqnarray}
\Cl = \left(\frac{\partial B}{\partial T}\right)^{-2} b(k)^2 \ \left [ 1 + 2 r \beta  \mu^2 + \beta^2 \mu^4 \right ]  D_{FoG} (k_{\parallel}, \sigma_p) P(k), 
\label{eq:HIPS}
\end{eqnarray}
This model includes the effect of scale-dependent complex bias $b(k), r$ and various the redshift space distortion effects $\mu, D_{FoG}$ \citep{1987MNRAS.227....1K, 2018MNRAS.476...96S, 2019JCAP...09..024M}. Here we assume standard $\Lambda$CDM cosmology with parameters taken from \citep{2016A&A...594A..13P} . The dark matter power spectrum is adopted from \citet{1994MNRAS.267.1020P}. 

\begin{table}
\begin{tabular}{|c|c|c|c|c|}
\hline
 redshift & z=1.25 & z= 1.5 & z=2.5 \\

\hline
$\theta_0$ ($^{'}$) & $22$ & $24$ & $39$ \\

\hline
$U_M$ [k$\lambda$]  & $52.6$ & $47.3$   & $29.6$ 
\\

\hline
\end{tabular}
\caption{The pamaters  $\theta_0$ and maximum baselines for observations of redshifted 21-cm emission  from redshifts 1.25, 1.5 and 3 for the uGMRT.}
\label{tab:GMRT}
\end{table}

\begin{table*}
\centering
\begin{tabular}{l|r|c|l|c|c|c|c|c|c}
\hline
Models  & $z_l$     &  $\theta_{x0}$ &  $\theta_{y0}$  & $\epsilon$ &  $\chi$  &  $\theta_a$  & $\theta_s$  & $\sigma_v$  &  References \\  &
             &  ($''$) &  ($''$) &             & ($^{\circ}$)& ($''$)  & ($''$)  & (km sec$^{-1}$)  &\\
\hline  
\hline
A1413  & 0.143  & 0  & 0 &  0.67  &  85.1  &  25.9  &  386.3  & 941  & 3\\ &  & 0   & 0  & 0.71  &  65.0 &  0.02  & 48.6  & 334 &  \\ & & 13.2 &-19.9 & 0.12  & 36.6  &  0.04  & 12.1  & 168 &    
\\
\hline
A2204 & 0.152  &0   & 0  & 0.54 & 134.6  & 4.9   & 366.5  & 556  & 3
\\
\hline
A868 & 0.153  & 0  &  0  &  0  & -66.5  &  26.0   &  364.5  & 1078  & 3 \\&  & -21.5  &  11.7  &  0.42  &  26.2  &  22.9  &  364.5  &  426  &   
\\
\hline
A2218   &0.171   & 3.1 & 20.8  &  0.04  &  38.0  &19.4 &198.4 &  697  &  3 \\ &  & -16.9  &  -21.7  & 0.32  & 9.2 &  39.8 & 161 &  992 &      \\&   & -0.5 & 0.1 & 0.46  & 52.4  & 1.7 & 12.7  & 506 &  \\
\hline
A1689  & 0.183    & 0.6   & -8.9       & 0.21       & 91.8         & 32      & 477.5    & 1437  & 6        \\
        &     & -70.0    & 47.8       & 0.80        & 80.5        & 22.1      & 157.8    & 643    & 
\\
\hline
A383 &0.187 & -0.3  & 0.5  & 0.15  & 123.7 & 88.3  & 309.6  & 1976  &  3
\\
\hline
A209  &0.200  & 0  & 0  & 0.57  & 43.0  &  14.7  & 293.7  &  630 & 1
\\
\hline
A963 &0.206  & 0   &  0   & 0.21  & 85.0  &  6.7  &  286.9  & 74 & 3
\\
\hline
A773 & 0.217 & 0  & 0  & 0.62  &  -37.3  & 11.6  &  275.7   &  501 & 3 \\ &  & 0  & 24  & 0.47  & -20.2  &  35.3  & 275.7  & 836   & \\&  & -119 &  6$^+$& 0.42  & -54.4$^+$ & 20.7 &   275.7 &  996 & 
\\
\hline
A2219  &0.228 & 0.1  & 0.2 & 0.65 & 32.9  &  20.4 &  265.5 & 854 & 3 \\
&   & -39.2 & -32 &  0.10 & 7.6  & 41.7  &  265.5 & 781  & \\
&    & -22.9  & 4.5 & 0.00 & 0.0 & 2.1 & 265.5  &  328    & 
\\
\hline
A267 & 0.230   & 0   &  0   & 0.6  &  -60.0  & 30.3  & 263.8  & 1060  & 1
\\
\hline
A2667  & 0.230  &0.1$^+$  & -0.5$^+$  &  0.32   &  -44.1  &  21.8  & 342.4  & 1114 & 3
\\
\hline
A2390  & 0.231    & 38.9 &  27.4  & 0.61  &  215.1  & 155.7 & 525.8 & 2038 & 3 \\&   & -0.9  & -1.4 & 0.03$^+$ & 30.5  &  7.86  & 77.5  &  633   & \\&   & 46.9  & 12.8  & 0.35  & 143.7  & 0.01  & 10.9 & 152  &  
\\
\hline
A521 & 0.247 & 0  &  0  & 0.67  & 49.5  & 4.6   & 250.1  &  553 & 3
\\
\hline
A1835 & 0.253 & 4.8 & 0.7  &  0.57  &  77.7  &  24.4  &  245.7  & 1219 & 3
\\
\hline

A68 &0.255  &  -1.5 & 0.2$^+$   &  0.53  &  125.9  & 21.5  & 302.7$^+$  &  908  & 2 \\ &  & -48.4  &  63.2  & 0  & 0  & 15.9  &  329.8  & 757   & 
\\
\hline
A611 &0.288  & 0  &  0  & 0.37  & -47.3  &  9.5  & 223.9 & 854  & 3
\\
\hline
A2744  & 0.308    & -4.9   & 2.7       & 0.28       & 65         & 45.8  & 213.8    & 1263   &5        \\ 
        &     & -15.7    & -17.2       & 0.61       & 43.3         & 9.3      &213.8   & 134       &
\\
\hline
A370  & 0.375  & 0.1 & 34.9 &  0.1  & 89.4  & 17.8  & 281.8 & 1040  & 4 \\ &   & -2.5 & 1.9  &  0.47  &  80.8  &  16.6  &  281.8  & 969    &  
\\
\hline
\end{tabular}
\caption{PIEMD model parameters of the projected lensing potential for clusters in our sample.  Columns are having values of redshift, centre of the ellipsoid, ellipticity, orientation, core and cut
radii,  velocity dispersion of these mass components. The '+' against some of the parameters represents those having less than one-sigma significance. The models are taken from 1. \cite{2005MNRAS.359..417S}
2. \cite{2007ApJ...662..781R}
3. \cite{2010MNRAS.404..325R}
4. \cite{2014ApJ...797...48J}
5. \cite{2015MNRAS.452.1437J} 6. \cite{2016A&A...590A..14B} }
\label{tab:LensMod}
\end{table*}

\begin{figure}
   \vspace{-0.5cm}
    \hspace{-0.7cm}
    \includegraphics[width=9cm,height=6cm]{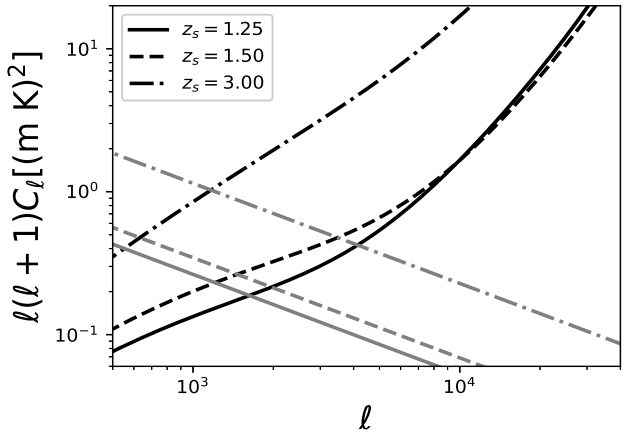}
    \caption{ Black curves show 21-cm angular power spectra for redshifts of 1.25, 1.5, 3.0 using the models from  \citet{2005MNRAS.356.1519B, 2016MNRAS.460.4310S, 2018MNRAS.476...96S,2016A&A...594A..13P} as described in the text. The grey line shows the contribution from the Diffused Galactic Synchrotron Emission (DGSE) at the corresponding  redshifted \HI frequencies. The DGSE is multiplied by $10^6$ for display purposes.}
    \label{fig:Cl}
\end{figure}

It is well known that the luminosity distances in the standard cosmology increase monotonically with redshift, however, the angular diameter distance increase to a redshift of $\sim 1.5$ and then decrease slowly. We choose to look at the 21-cm angular power spectrum from redshifts of $(1.25, 1.5, 3.0)$. This lets us access the effect of the lensed 21-cm angular power spectrum estimator  against the  change of angular diameter and luminosity distance with redshift. For the observational purpose, we consider the parameters and baseline of the uGMRT\footnote{uGMRT: upgraded Giant Meterwave Radio Telescope} \citep{2017CSci..113..707G}. The uGMRT has an aperture diameter of 45 meters with the largest antenna separation of 25 km. For simplicity, here we assume a Gaussian aperture for the antenna and list the values of the parameter $\theta_0$ corresponding for the three redshifts in Table~\ref{tab:GMRT}. The maximum antenna separation for the uGMRT  is $25$ km. Since the angular multipole $l = 2\, \pi \, U$, the range of multipoles at which the uGMRT is sensitive to are $\sim310-330,000$, $\sim280 -300,000$, $\sim180-187,000$ for the redshifts $1.25, 1.5$ and $3.0$ respectively. In this work, we plan to combine the power spectrum of redshifted 21-cm emission behind different clusters.
Typically a galaxy cluster is about one Mpc across. Combining  the \HI power spectrum at scales smaller than the typical size of a cluster from different directions may introduce a large number of local fluctuations  and hence the combined lens estimator would not be much effective.
Furthermore, the baseline coverage of the uGMRT at larger baselines becomes sparse and hence signal to noise is expected to decrease. Using models of gravitational lenses, \citet{2020MNRAS.498.3275A} have shown that the lensing sampling function falls drastically beyond a certain baseline, and would not be able to suppress noise effectively with the estimator discussed here. We restrict our investigation to an angular multipole range of $1000 - 20,000$ in this work, which corresponds to $\sim 0.5$ Mpc at the source redshifts.  Figure~\ref{fig:Cl} shows the expected 21-cm angular power spectrum  at the the redshifted of $1.25, 1.5$ and $3.0$.  We also show the power spectra of the Diffused Galactic Synchrotron Radiation  (DGSE) in the same plot with grey lines at the three redshifted frequencies of \HI. Note that the DGSE values are multiplied by $10^{-6}$ for display purposes.

\subsection{Lens Models}
Several methods have been devised to estimate the lensing potentials using the optical study of strong lensing by galaxy clusters \citep{2007NJPh....9..447J,2010MNRAS.402L..44R,2016A&A...590A..14B,2018ApJ...859..159C,2018MNRAS.481.1427S}. Most of the lensing models achieved in this way are parametric. The parametric forms include NFW profile \citep{2000PASJ...52...99A,1998ApJ...502..531B}, Soften Power-law Elliptical Potential \citep{1998ApJ...502..531B}, singular isothermal sphere \citep{1998ApJ...502..531B, 2002ApJ...566..652L} etc. It is observed that in a particular lens, several dark halos of the above types can be present. In addition to these, the lensing potential also has contribution from the individual galaxy masses in the cluster. However, the large dark matter halos are the ones, which contribute most to the  overall strong lensing \citep{1998ApJ...502..531B}. The widely used parametric model for lensing potential is the pseudo isothermal elliptical mass distribution (henceforth PIEMD) model \citet{1993ApJ...417..450K}. In this model the projected potential of each halo  in  the lens
is described in terms of the ellipticity $\epsilon$, ellipse position angle $\chi$, two characteristic radii $\theta_a, \theta_s$ and effective dispersion velocity $\sigma_v$. 
Since the angular extent of the lenses is rather small, the lensing potential is given as a function of the angular coordinates $(\theta_x, \theta_y)$, the  position in the lens as a cartesian component of the angular separation in the sky plane with respect to the reference centre of the component $(\theta_{x0}, \theta_{y0})$. Note that, in case of multiple PIEMD components, the component centre $(\theta_{x0}, \theta_{y0})$ can vary.
A particular galaxy cluster lens can have one or more PIEMD components. The projected gravitational potential  can be written as 
\begin{eqnarray}
\psi(\theta_x, \theta_y) &=& 6\pi \frac{D_{ds}}{D_s} \frac{\theta_s + \theta_a}{\theta_s} \frac{\sigma_{v}^2}{c^2} \ f(\theta),  \\ \nonumber 
f(\theta) &=&   \sqrt{\theta_s^2+\theta^2}-\sqrt{\theta_a^2+\theta^2} \\ \nonumber
	 &+& \theta_a \ln \left( \theta_a+\sqrt{\theta_a^2+\theta^2} \right) -\theta_s \ln \left (\theta_s+\sqrt{\theta_s^2+\theta^2} \right),
\end{eqnarray}    
Here, $D_{ds}$ and $D_s$ are the  angular diameter distances between lens-source and the source-observers respectively.  The variable $\theta$ is given as 
\begin{eqnarray}
\theta^2 &=&\left[ \frac{ (\theta_x - \theta_{x0})\cos\chi +(\theta_y - \theta_{y0})\sin \chi }{1+\epsilon}\right]^2 \\ \nonumber &+ & \left [ \frac{-(\theta_x-\theta_{x0})\sin \chi+(\theta_y-\theta_{y0})\cos \chi}{1-\epsilon} \right] ^2.
\end{eqnarray}

\begin{figure}
    \begin{center}
    \includegraphics[scale=0.5]{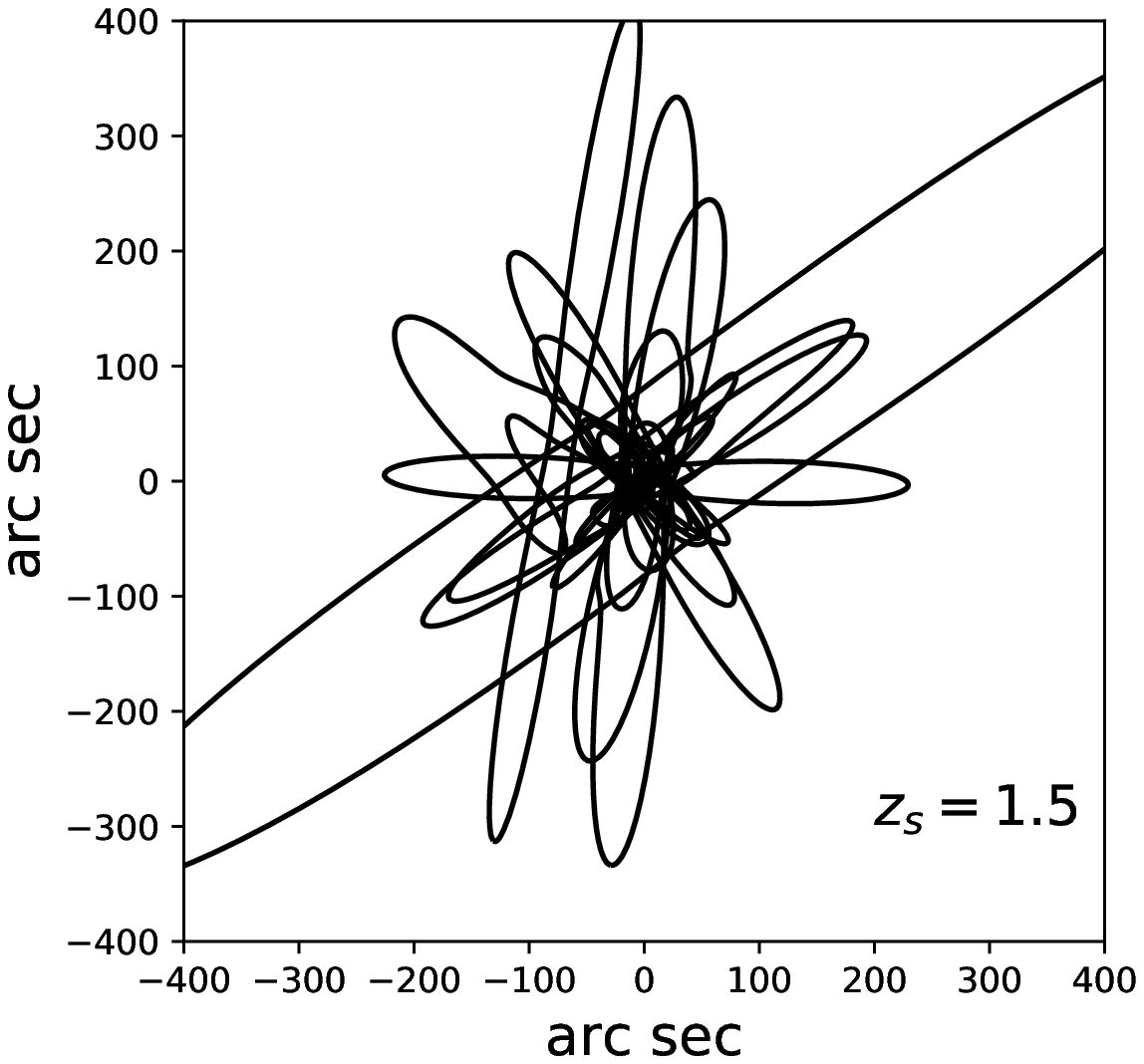}
    \caption{ Central part of the critical curves for all the lens models given in Table~\ref{tab:LensMod} shown for a source redshift of 1.5.}
    \label{fig:critical}
    \end{center}
\end{figure}

Here we consider the contributions from all the components with  $\theta_s >10^{''}$ for the calculations. Table~\ref{tab:LensMod} gives the lens parameters for $19$ clusters from the Abell  cluster catalogue.  The clusters are arranged in increasing order of their redshifts.  The lens models used here are obtained from the references given in the table, where they use lensed multiple images of background sources in optical wavelengths to model the cluster potential. For some of the clusters, a few parameters are not well constrained as of now, these are marked with a star '+' against their names. We also restrict to clusters with at least most of the parameters well constrained and not considered clusters with a large elliptic component. Figure~\ref{fig:critical} shows the critical curves for all the cluster lenses if they would have been in the same direction in the sky. The source redshift is taken to be 1.5. This show that using multiple lenses we can obtain magnification over a large fraction of the sky. We also notice that the sampling of the sky is better at smaller angular scales.

\begin{figure}
    \begin{center}
    \includegraphics[scale=0.5]{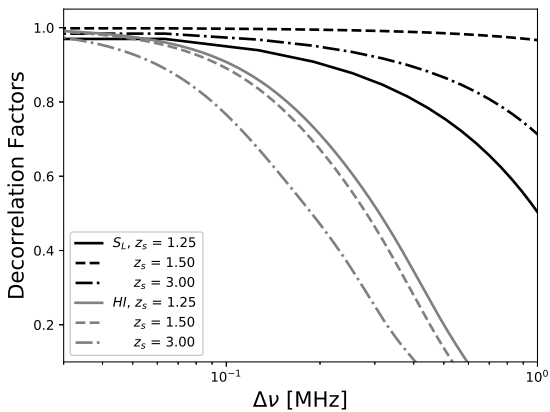}
    \caption{ Figure showing how the lensing sampling function at the multipole of 1000 remains correlated with frequency. Here y-axis is in a relative scale with the value at $\Delta \nu$ set to unity. The black lines correspond to the lensing sampling function and the grey lines correspond to the \HI signal.}
    \label{fig:Slfreq}
    \end{center}
\end{figure}

\subsection{Results}
In radio interferometric observations, noise in each observed visibility $\sigma_N$ observed in a given frequency channel depends on the  source equivalent flux density (SEFD) of the observing telescope, the integration time of observation $\Delta \tau$ and the width of the frequency channel over which the signal is integrated with frequency. If a signal remains correlated over a certain frequency range $\nu_c$, then one can choose that as the width of the frequency channel of observation, provided bandwidth smearing effect is not important. Assuming that the property of the signal does not change over nearby channels of width $\nu_c$, we can estimate independent samples of the signal by including several channels to a total bandwidth of  $B_w$. Then effectively,  
\begin{equation}
\label{eq:fsep}
\sigma^2_N = \frac{[SEFD]^2}{\Delta \tau \sqrt{B_w \nu_c}}.
\end{equation}
\begin{figure*}
    \begin{center}
    \includegraphics[scale=0.6]{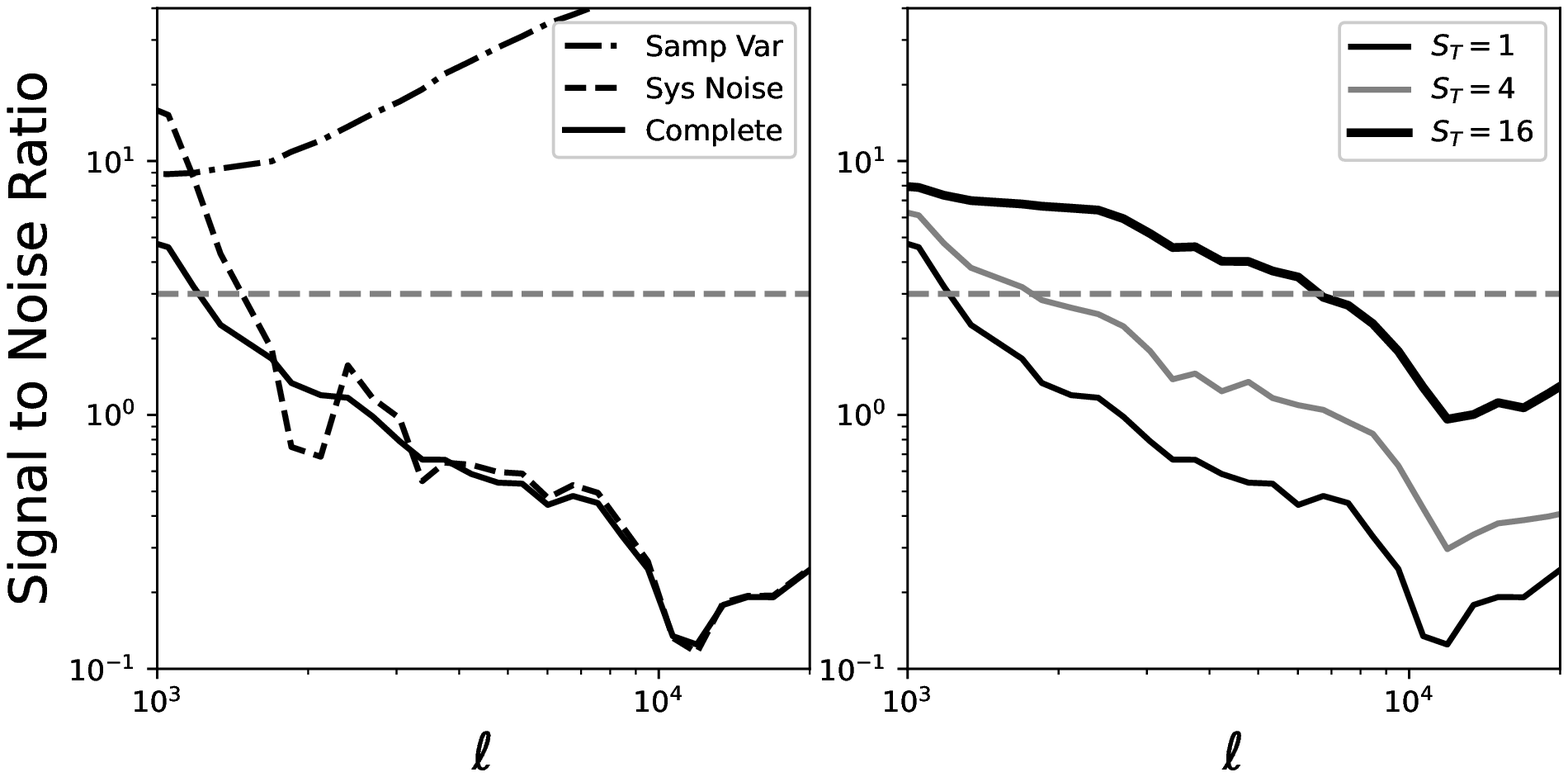}
    \caption{Signal to noise ratio (SNR) for the cluster Abell~773 for $250$ hours observations with the uGMRT with a bandwidth of $16$ MHz. Left panel: The black continuous line shows the signal to noise ratio for $S_T=1$. The dashed line corresponds to the case when there is no effect of sample variance, whereas the dot-dashed line shows the SNR if instrumental noise is negligible. Right panel: Signal to noise ratio for different values of $S_T$.  In both panels, the grey dashed line marks three-sigma confidence level.}
    \label{fig:STSNR}
    \end{center}
\end{figure*}
\begin{figure*}
    \begin{center}
    \includegraphics[scale=0.85]{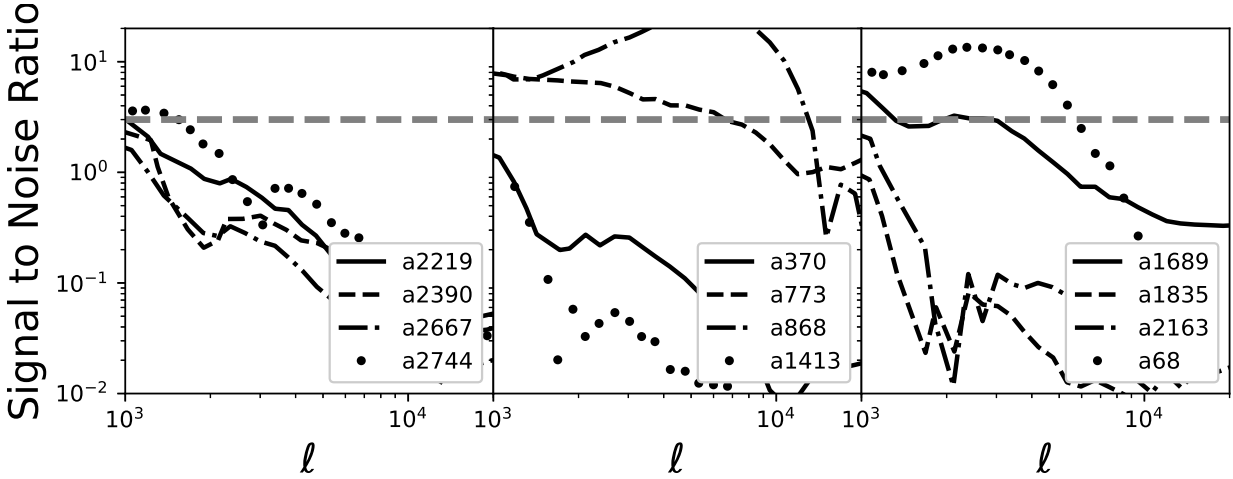}
    \caption{ Signal to noise ratio for the best clusters in our sample for $250$ hours observations with the uGMRT with a bandwidth of $16$ MHz shown in three panels. The grey dashed line marks three-sigma confidence level.}
    \label{fig:AllClust}
    \end{center}
\end{figure*}

\begin{figure*}
    \begin{center}
    \includegraphics[scale=0.6]{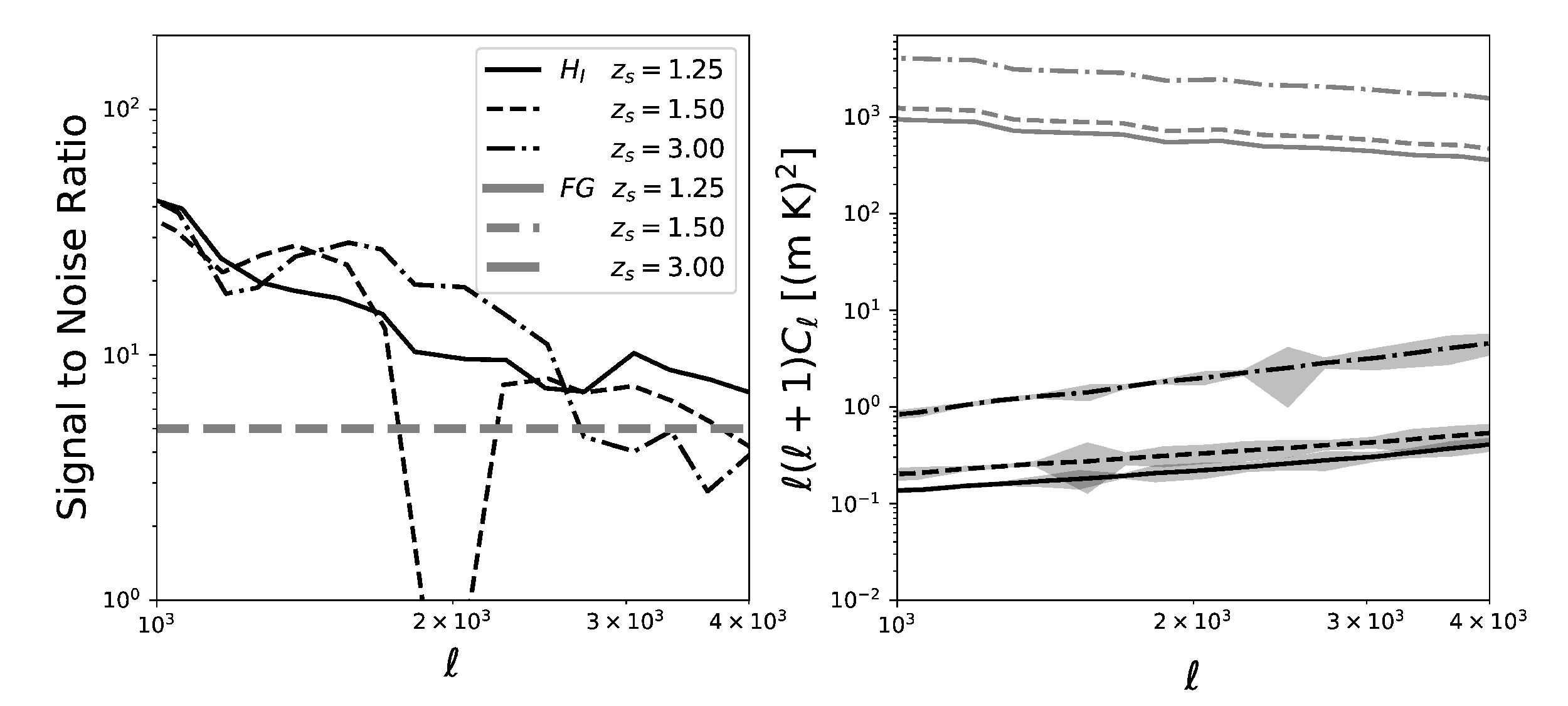}
    \caption{  Left panel: Signal to Noise ratio combining eight best  clusters in our sample for redshifts of $1.25, 1.5$ and $3.0$.  Each case  is for $16$ MHz bandwidth uGMRT observation. For redshifts  of $1.25$ and $1.5$ with the total observation hours for all eight clusters is $400$ hours, for the redshift of $1.5$ the observation time is $200$ hours only. The grey line shows the five-sigma confidence level. Right panel: Angular power spectrum at redshifts z= 1.25, 1.5 and 3.0 and corresponding modified DGSE. The grey region show the one-sigma error in the power spectrum estimates. Legends for different cases are shown in the left panel and is used identically for both the panels of this figure.}
    \label{fig:Comb}
    \end{center}
\end{figure*}

The observed interferometric signal we discuss here depends on the sky brightness distribution at the source redshift as well as the property of the gravitational lens. \citet{2005MNRAS.356.1519B} has discussed how the 21-cm signal decorrelates with angular frequencies. They show that at different redshifts and angular multipoles, the bandwidth over which the signal remain correlated changes. Using their result, we find that the visibility correlation signal remains correlated for a frequency separation of $350$ kHz, $400$  KHz and $4.1$ MHz at redshifts of $1.25, 1.5$ and $3.0$ respectively at the angular multipole of $1000$. Here, in addition, we need to investigate how the lensing sampling function changes with frequency. Note that, change in the lensing sampling function with frequency is expected to depend on the particular lens used. Figure~\ref{fig:Slfreq} shows quantity $\langle S_L^{*}(\nu) S_L(\nu + \Delta \nu) \rangle/ \langle \mid S_L^{*}(\nu)\mid^2  \rangle$ as a function of frequency separation $\Delta \nu$ for these redshifts at the angular multipole of $1000$ for the lens Abell~773. The de-correlation of the lensing sampling function is slower than that of the sky visibility correlation. Hence, we choose the de-correlation frequency of the sky visibility correlation as the value for $\nu_c$. We shall mention any change in these parameters whenever necessary. We choose $N_T = 5$ for the estimations henceforth, which gives at least $10$ baseline pairs within a grid. We also neglect the effect of foreground here, its effect will be discussed later.

We first show the signal to noise ratio achievable with $250$ hours of the uGMRT observations of 21-cm signal from a redshift of $1.5$ for the cluster lens Abel~773 with a $16$ MHz of bandwidth. This cluster is a rather nearby cluster at a redshift of $0.217$. The left panel of Figure~\ref{fig:STSNR} show the contribution to signal to noise ratio arising if the system noise is negligible (dash-dot), sample variance is negligible (dashed) and if all the contributions to noise are present (solid). The grey dashed horizontal line indicate a signal to noise of three. All these curves are shown for $S_T = 1$. The right panel of the same figure demonstrate the variation of the signal to noise ratio as threshold value $S_T$ is increased. We observe that at a $S_T$ of $16$, the signal to noise ratio for the angular power spectrum for $250$ hours of the uGMRT observation with a bandwidth of $16$ MHz through the cluster lens Abel~773 is above three-sigma till an angular multipole of $\sim 6000$.  We observe that the SNR is lower for the low angular multipoles as a result of the low number of sampling, whereas reducing the system noise is more important to achieve higher SNR at the larger multipoles.

Clearly, with $S_T = 16$, we recover the \HI signal significantly above the three-sigma confidence. We next use $S_T = 16$ to access the efficacy of the gravitational lenses for all the clusters in Table~\ref{tab:LensMod}.
Three panels of Figure~\ref{fig:AllClust} show signal to noise ratio for the clusters in our sample with values greater than unity anywhere for $l>1000$. The horizontal thick black dashed line corresponds to the three-sigma significance. Clearly, most of the clusters are not much effective as a lens to enhance the \HI signal significantly. The clusters, we find as good candidates for post-EoR \HI signal enhancement are: 
Abell~a68,
Abell~a868,
Abell~1689,
Abell~2219,
Abell~2390,
Abell~773,
Abell~2744,
Abell~370.

Figure~\ref{fig:STSNR} demonstrated that the signal to noise ratio in the \HI signal improves with $S_T = 16$. Increasing $S_T$ further reduces the number of grids available in the baseline plane and the sample variance error increase significantly. Subsequently, we used  $S_T=16$ for all the clusters in our sample to estimate the expected signal to noise ratio of lensed redshifted 21-cm signal from a redshift of $1.5$. We found that nine of the clusters in our sample reduce  the noise in the estimator  to three-sigma significance for a bandwidth of $16$ MHz and an observation time of $250$ hours. However, for individual clusters, the range of angular multipoles over which we are sensitive is limited to  certain angular multipole ranges. Next, we use the combined angular power spectrum estimates from all nine clusters. Since,  while using multiple clusters we have estimates of the angular power spectra in enough number of grids in the baseline plane, we choose a value of $S_T=32$. 

The left panel of Figure~\ref{fig:Comb} shows the signal to noise estimates of the lensed redshifted \HI angular power spectra (black curves) at the redshifts of $1.25, 1.5$ and $3.0$. We have chosen a bandwidth of $16$ MHz and a total observation time of $400$ hours here for the redshifts of $1.25$ and $1.5$. For the source redshift $3.0$, we have a higher signal to noise and the plot shown here is with a total observation time of $200$ hours.  The horizontal grey line indicates a signal to noise of $5$. We find that $400$ hours of observation is enough to observe lensed \HI signal from the redshifts of $1.25, 1.5$ and $200$ hours of observation is enough to observe the lensed \HI signal from the redshift of $3.0$ with  five-sigma significance.

\subsection{Effect of Foreground}
The sky visibilities $V_S$ in observation, at redshifted 21-cm frequencies, are dominated by the foreground  diffused  galactic synchrotron radiation from our galaxy. The angular power spectrum of this radiation is shown in Figure~\ref{fig:Cl} along with the redshifted 21-cm signal. To observe the redshifted 21-cm signal one need to model and subtract the diffused synchrotron foreground signal. Given that the amplitude of the foreground is much higher than the redshifted 21-cm signal, such observations require high dynamic range calibration of the observed visibilities \citep{2010ApJ...724..526D,2020MNRAS.495.3683K}. The lensed 21-cm angular power spectrum estimator reduces the effective amplitude of this foreground and hence improve the calibration requirements. 
Note that in presence of foreground, eqn.~\ref{eq:LS} would have an additional term, $V_{Fi}$ for the foreground, which is not enhanced by $S_L$. Since the foreground is not correlated with the 21-cm signal, the observed value of $E_L(g)$ in a given grid modifies to $C_{l_g} + C_{Fl_g} /\mid S_L(g)\mid^2$ approximately, where $C_{Fl_g} $ is the angular power spectrum of the foreground and $S_L(g)$ is the lensing sampling function in the grid $g$. Since, $S_T$ is chosen to be $32$ here, the effective contribution of the foreground in $E_L(g)$ decreases. 

Black lines in the right panel of Figure~\ref{fig:Comb} show the angular power spectra of the redshifted 21-cm signal at redshifts of $1.25, 1.5$ and $3.0$. We also show the one-sigma errors with the grey bands for all three redshifts.  The lens modified foreground angular power spectra corresponding to the \HI signal from the same redshifts are shown with grey lines. Comparing with figure \ref{fig:Cl} we observe that  the foreground is suppressed approximately by a factor of $1000$, however, is still significantly higher compared to the redshifted 21-cm signal. We believe this would help significantly to avoid excess bias and variance in the signal that may arise with a higher strength foreground through calibration errors. 

\section{Discussion and Conclusion}
\citet{2020MNRAS.498.3275A} demonstrated that the strong gravitational lensing by clusters has the potential to enhance the redshifted 21-cm power spectra behind them. In this work, we introduce a lensing based power spectrum estimator and calculate the uncertainties associated with these estimates. The estimator gives an unbiased estimate of the redshifted 21-cm power spectrum and suppresses the observational uncertainties. We use strong lens models of a few clusters known from optical studies to investigate the efficacy of this estimator in enhancing the 21-cm angular power spectra from the post reionization era from redshifts of $1.25, 1.5$ and 3.0. Since the strong lensing enhancement is different at different baselines, it depends on the quality of the baseline coverage of the radio interferometer with which the 21-cm signal is observed. We consider uGMRT baseline configuration in this work. We find that for eight  of the clusters in our sample, the lensed power spectrum estimates are above three-sigma uncertainty for a limited range of angular multipoles. Assuming statistical homogeneity, we showed that by combining the power spectrum estimates from these eight clusters it is possible to achieve statistically significant detection of post reionization 21-cm power spectrum using the lensed \HI power spectrum estimator.

The highest redshift  of a cluster in our sample is $0.375$.  We find that for $250$ hours of uGMRT observation with a $16$ MHz of bandwidth,   three clusters in our sample, Abell~773, Abell~68 and Abell~868  enhances the power spectra of the 21-cm emission from redshifts of 1.25, 1.5 and 3.0 statistically significantly. It may be possible to detect \HI power spectrum using any of these clusters only (see further discussions below). These clusters are at redshifts of $0.217$, $0.255$ and $0.153$ respectively. We also find that these clusters effectively suppress the noise for angular multipoles $l<6000$.  Five more clusters show significant enhancement of the signal to noise ratio of the estimator, however, with the similar time and bandwidth, they individually are not effective to estimate the redshifted \HI signal. We assume statistical homogeneity and combine the angular power spectrum estimates from the direction of these eight clusters. It is found that a total of $400$ hours of the uGMRT observation of these eight clusters, each with a bandwidth of $16$ MHz, would let us estimate the 21-cm power spectra of \HI at redshifts of $1.25$ and $1.5$ with five-sigma confidence up to an angular multipole of $4000$. A similar sensitivity at a similar angular multipole range can be achieved with a total of $200$ hours of observations of these clusters with the uGMRT for a redshift of 3.0. 

The lens power spectrum estimator discussed here does not only suppress the uncertainties in the estimates but also reduce the foreground contribution from the diffused galactic synchrotron radiation. This is because,  later is not modified by the lensing potential, unlike the 21-cm signal. We found that for the parameters of the estimator chosen here, the effective foreground power spectra is reduced by a factor of $1000$. It is to note that  this does not suppress the foreground below the 21-cm power spectrum. \citet{2020MNRAS.495.3683K} demonstrated that the residual grain errors in 21-cm observations in presence of strong foreground is a major challenge for reionization and post reionization power spectrum measurements. The residual gain errors through the strong foreground signal introduce a bias in the 21-cm power spectrum estimates and enhance its variance. The three orders of magnitude suppression of the foreground signal by the lens power spectrum estimator discussed here are expected to  reduce both the bias and variance of the signal arising through the residual gain errors significantly.

\citet{2010MNRAS.407..567B} showed that with a $16$ MHz of bandwidth with the uGMRT baseline configuration  a $1000$ hours of observation can detect the redshifted 21-cm signal over the noise for a baseline $<600 \ \lambda$, i.e, angular multipoles lesser than $3800$ for a redshift of $1.33$. They show that at a redshift of $3.3$, the detection for GMRT is possible with three-sigma for about $2000$ hours of observations. Thanks to the strong lensing by the clusters, the lensed visibility correlation estimator discussed here requires lesser observation time with the same observational bandwidth. Furthermore, the estimator discussed here effectively reduce the foreground contamination in the signal suppressing the systematics.

A major assumption in this work is that for the gravitational lenses we have well established parametric models of their lensing potential. This is far from the truth at present, uncertainties of some of the parameters in Table~\ref{tab:LensMod} are indicated by the "+" sign. Furthermore, one needs to keep in mind that the lens models available today have different uncertainties at the different physical positions of the lensing potential \citep{2014MNRAS.444..268R}, which may lead to additional complications. 
We expect with the advent of optical studies of these lenses, the lensing models would become more precise. Considering eqn.~(\ref{eq:est}), the effect of lensing is carried by the weight factor $w_{ij}$ through $k_{ij} = (S_{Li} S^{*}_{Lj})^{-1}$, that is the power spectrum estimates depend on the lens sampling function quadratically.  If we assume that there is a $10 \%$ error in the lensing magnification estimates, then this would lead to an additional  $20 \%$  uncertainties in the power spectrum estimates through the estimator discussed here. Hence, for detection of the 21-cm power spectrum  with five-sigma significance, through the lensed visibility correlation estimator discussed here,  we need to know the lensing magnification to an accuracy of $10 \%$ or lower. Note that, though the efficacy of this estimator is limited to the accuracy of the lens models, it is not limited to a particular type of lens model, like the PIEMD model discussed here. If any good estimates of the lens model are available, they can be incorporated into this estimator. 

The galaxy clusters we use here as gravitational lenses also have emissions in the radio continuum and that  is expected to contaminate   the lensed redshifted 21-cm radiation. We have not included this in our foreground estimates here.  The continuum signal from the cluster is smooth over a rather large frequency range and   a continuum subtraction based method  can be used to mitigate it. 

In summary, in this work, we introduce a strong lensing based visibility correlation estimator and the uncertainties related to it. We demonstrated with some of the best-known lens models that by combining the power spectrum estimates from several lenses the uGMRT observations of a few hundred hours would be good to estimate the redshifted 21-cm power spectrum given that the lens models are well constrained. The estimator also reduces the effect of contamination from the diffused galactic foreground emission and requirement of high dynamic range calibration accuracy. We show, with a rough estimate that the lens models need to be known to an accuracy such that the magnifications can be estimated to a  $10 \%$ uncertainty  for the detection of the 21-cm signal at five-sigma significance. Our result suggests that  a way forward would be to  identify a few of these cluster lenses and establish their lens models. We hope that as the lens models become better in future the method presented in this work would become an effective and efficient tool to estimate the post reionization 21-cm power spectra.

\section*{Acknowledgement}
PD and UA acknowledge useful discussion with Tapamoy Guha Sarkar. UA acknowledge GATE fellowship for funding this work.
UA is thankful to Meera Nandakumar, Jais Kumar and Pavan Kumar Vishwakarma for useful discussions during this work.

\section*{DATA AVAILABILITY}
No new data were generated or analysed in support of this research.
\newcommand{\newblock}{}
\bibliographystyle{mn2e}
\bibliography{reference}

\begin{thebibliography}{}

\bibitem[\protect\citeauthoryear{{Ade et al.} \& Collaboration}{{Ade et al.} \&
  Collaboration}{2016}]{2016A&A...594A..13P}
{Ade et al.} Collaboration P.,  2016, \aap, 594, A13

\bibitem[\protect\citeauthoryear{{Ali} \& {Bharadwaj}}{{Ali} \&
  {Bharadwaj}}{2014}]{2014JApA...35..157A}
{Ali} S.~S.,  {Bharadwaj} S.,  2014, Journal of Astrophysics and Astronomy, 35,
  157

\bibitem[\protect\citeauthoryear{{Arora} \& {Dutta}}{{Arora} \&
  {Dutta}}{2020}]{2020MNRAS.498.3275A}
{Arora} U.,  {Dutta} P.,  2020, \mnras, 498, 3275

\bibitem[\protect\citeauthoryear{{Asano}}{{Asano}}{2000}]{2000PASJ...52...99A}
{Asano} K.,  2000, \pasj, 52, 99

\bibitem[\protect\citeauthoryear{{Bagla}, {Khandai} \& {Datta}}{{Bagla}
  et~al.}{2010}]{2010MNRAS.407..567B}
{Bagla} J.~S.,  {Khandai} N.,    {Datta} K.~K.,  2010, \mnras, 407, 567

\bibitem[\protect\citeauthoryear{{Barkana}}{{Barkana}}{1998}]{1998ApJ...502..531B}
{Barkana} R.,  1998, \apj, 502, 531

\bibitem[\protect\citeauthoryear{{Barkana} \& {Loeb}}{{Barkana} \&
  {Loeb}}{2001}]{2001PhR...349..125B}
{Barkana} R.,  {Loeb} A.,  2001, \physrep, 349, 125

\bibitem[\protect\citeauthoryear{{Baugh}, {Cole}, {Frenk} \& {Lacey}}{{Baugh}
  et~al.}{1998}]{1998ApJ...498..504B}
{Baugh} C.~M.,  {Cole} S.,  {Frenk} C.~S.,    {Lacey} C.~G.,  1998, \apj, 498,
  504

\bibitem[\protect\citeauthoryear{{Bera}, {Kanekar}, {Chengalur} \&
  {Bagla}}{{Bera} et~al.}{2019}]{2019ApJ...882L...7B}
{Bera} A.,  {Kanekar} N.,  {Chengalur} J.~N.,    {Bagla} J.~S.,  2019, \apjl,
  882, L7

\bibitem[\protect\citeauthoryear{{Bharadwaj} \& {Ali}}{{Bharadwaj} \&
  {Ali}}{2005}]{2005MNRAS.356.1519B}
{Bharadwaj} S.,  {Ali} S.~S.,  2005, \mnras, 356, 1519

\bibitem[\protect\citeauthoryear{{Bharadwaj}, {Nath} \& {Sethi}}{{Bharadwaj}
  et~al.}{2001}]{2001JApA...22...21B}
{Bharadwaj} S.,  {Nath} B.~B.,    {Sethi} S.~K.,  2001, Journal of Astrophysics
  and Astronomy, 22, 21

\bibitem[\protect\citeauthoryear{{Bina}, {Pell{\'o}}, {Richard}, {Lewis},
  {Patr{\'\i}cio}, {Cantalupo}, {Herenz}, {Soto}, {Weilbacher}, {Bacon},
  {Vernet}, {Wisotzki}, {Cl{\'e}ment}, {Cuby}, {Lagattuta}, {Soucail} \&
  {Verhamme}}{{Bina} et~al.}{2016}]{2016A&A...590A..14B}
{Bina} D.,  {Pell{\'o}} R.,  {Richard} J.,  {Lewis} J.,  {Patr{\'\i}cio} V.,
  {Cantalupo} S.,  {Herenz} E.~C.,  {Soto} K.,  {Weilbacher} P.~M.,  {Bacon}
  R.,  {Vernet} J.~D.~R.,  {Wisotzki} L.,  {Cl{\'e}ment} B.,  {Cuby} J.~G.,
  {Lagattuta} D.~J.,  {Soucail} G.,    {Verhamme} A.,  2016, \aap, 590, A14

\bibitem[\protect\citeauthoryear{{Blecher}, {Deane}, {Heywood} \&
  {Obreschkow}}{{Blecher} et~al.}{2019}]{2019MNRAS.484.3681B}
{Blecher} T.,  {Deane} R.,  {Heywood} I.,    {Obreschkow} D.,  2019, \mnras,
  484, 3681

\bibitem[\protect\citeauthoryear{{Bothwell}, {Kennicutt} \& {Lee}}{{Bothwell}
  et~al.}{2009}]{2009MNRAS.400..154B}
{Bothwell} M.~S.,  {Kennicutt} R.~C.,    {Lee} J.~C.,  2009, \mnras, 400, 154

\bibitem[\protect\citeauthoryear{{Bowman}, {Morales} \& {Hewitt}}{{Bowman}
  et~al.}{2009}]{2009ApJ...695..183B}
{Bowman} J.~D.,  {Morales} M.~F.,    {Hewitt} J.~N.,  2009, \apj, 695, 183

\bibitem[\protect\citeauthoryear{{Carucci}, {Villaescusa-Navarro} \&
  {Viel}}{{Carucci} et~al.}{2017}]{2017JCAP...04..001C}
{Carucci} I.~P.,  {Villaescusa-Navarro} F.,    {Viel} M.,  2017, \jcap, 2017,
  001

\bibitem[\protect\citeauthoryear{{Cerny}, {Sharon}, {Andrade-Santos}, {Avila},
  {Brada{\v{c}}}, {Bradley}, {Carrasco}, {Coe}, {Czakon}, {Dawson}, {Frye},
  {Hoag}, {Huang}, {Johnson}, {Jones}, {Lam}, {Lovisari}, {Mainali}, {Oesch},
  {Ogaz}, {Past}, {Paterno-Mahler}, {Peterson}, {Riess}, {Rodney}, {Ryan},
  {Salmon}, {Sendra-Server}, {Stark}, {Strolger}, {Trenti}, {Umetsu}, {Vulcani}
  \& {Zitrin}}{{Cerny} et~al.}{2018}]{2018ApJ...859..159C}
{Cerny} C.,  {Sharon} K.,  {Andrade-Santos} F.,  {Avila} R.~J.,  {Brada{\v{c}}}
  M.,  {Bradley} L.~D.,  {Carrasco} D.,  {Coe} D.,  {Czakon} N.~G.,  {Dawson}
  W.~A.,  {Frye} B.~L.,  {Hoag} A.,  {Huang} K.-H.,  {Johnson} T.~L.,  {Jones}
  C.,  {Lam} D.,  {Lovisari} L.,  {Mainali} R.,  {Oesch} P.~A.,  {Ogaz} S.,
  {Past} M.,  {Paterno-Mahler} R.,  {Peterson} A.,  {Riess} A.~G.,  {Rodney}
  S.~A.,  {Ryan} R.~E.,  {Salmon} B.,  {Sendra-Server} I.,  {Stark} D.~P.,
  {Strolger} L.-G.,  {Trenti} M.,  {Umetsu} K.,  {Vulcani} B.,    {Zitrin} A.,
  2018, \apj, 859, 159

\bibitem[\protect\citeauthoryear{{Chakraborty}, {Datta}, {Choudhuri}, {Roy},
  {Intema}, {Choudhury}, {Datta}, {Pal}, {Bharadwaj}, {Dutta} \&
  {Choudhury}}{{Chakraborty} et~al.}{2019}]{2019MNRAS.487.4102C}
{Chakraborty} A.,  {Datta} A.,  {Choudhuri} S.,  {Roy} N.,  {Intema} H.,
  {Choudhury} M.,  {Datta} K.~K.,  {Pal} S.,  {Bharadwaj} S.,  {Dutta} P.,
  {Choudhury} T.~R.,  2019, \mnras, 487, 4102

\bibitem[\protect\citeauthoryear{Chirivì, Yıldırım, Suyu \&
  Halkola}{Chirivì et~al.}{2020}]{Chiriv__2020}
Chirivì G.,  Yıldırım A.,  Suyu S.~H.,    Halkola A.,  2020, Astronomy \&
  Astrophysics, 643, A135

\bibitem[\protect\citeauthoryear{{Cho}, {Lazarian} \& {Timbie}}{{Cho}
  et~al.}{2012}]{2012ApJ...749..164C}
{Cho} J.,  {Lazarian} A.,    {Timbie} P.~T.,  2012, \apj, 749, 164

\bibitem[\protect\citeauthoryear{{Choudhuri}, {Bharadwaj}, {Chatterjee}, {Ali},
  {Roy} \& {Ghosh}}{{Choudhuri} et~al.}{2016}]{2016MNRAS.463.4093C}
{Choudhuri} S.,  {Bharadwaj} S.,  {Chatterjee} S.,  {Ali} S.~S.,  {Roy} N.,
  {Ghosh} A.,  2016, \mnras, 463, 4093

\bibitem[\protect\citeauthoryear{{Choudhuri}, {Bharadwaj}, {Ghosh} \&
  {Ali}}{{Choudhuri} et~al.}{2014}]{2014MNRAS.445.4351C}
{Choudhuri} S.,  {Bharadwaj} S.,  {Ghosh} A.,    {Ali} S.~S.,  2014, \mnras,
  445, 4351

\bibitem[\protect\citeauthoryear{{Chowdhury}, {Kanekar}, {Chengalur}, {Sethi}
  \& {Dwarakanath}}{{Chowdhury} et~al.}{2020}]{2020Natur.586..369C}
{Chowdhury} A.,  {Kanekar} N.,  {Chengalur} J.~N.,  {Sethi} S.,
  {Dwarakanath} K.~S.,  2020, \nat, 586, 369

\bibitem[\protect\citeauthoryear{{Datta}, {Bowman} \& {Carilli}}{{Datta}
  et~al.}{2010}]{2010ApJ...724..526D}
{Datta} A.,  {Bowman} J.~D.,    {Carilli} C.~L.,  2010, \apj, 724, 526

\bibitem[\protect\citeauthoryear{{Deane}, {Obreschkow} \& {Heywood}}{{Deane}
  et~al.}{2015}]{2015MNRAS.452L..49D}
{Deane} R.~P.,  {Obreschkow} D.,    {Heywood} I.,  2015, \mnras, 452, L49

\bibitem[\protect\citeauthoryear{{DeBoer}}{{DeBoer}}{2017}]{2017PASP..129d5001D}
{DeBoer} D. R. e.~a.,  2017, \pasp, 129, 045001

\bibitem[\protect\citeauthoryear{Dewdney, Braun \& Turner}{Dewdney
  et~al.}{2017}]{8105425}
Dewdney P.~E.,  Braun R.,    Turner W.,  2017, in 2017 XXXIInd General Assembly
  and Scientific Symposium of the International Union of Radio Science (URSI
  GASS) The mid-frequency telescope for the square kilometre array (ska-mid).
pp~1--4

\bibitem[\protect\citeauthoryear{{Dye}, {Furlanetto}, {Dunne}, {Eales},
  {Negrello}, {Nayyeri}, {van der Werf}, {Serjeant}, {Farrah},
  {Micha{\l}owski}, {Baes}, {Marchetti}, {Cooray}, {Riechers} \&
  {Amvrosiadis}}{{Dye} et~al.}{2018}]{2018MNRAS.476.4383D}
{Dye} S.,  {Furlanetto} C.,  {Dunne} L.,  {Eales} S.~A.,  {Negrello} M.,
  {Nayyeri} H.,  {van der Werf} P.~P.,  {Serjeant} S.,  {Farrah} D.,
  {Micha{\l}owski} M.~J.,  {Baes} M.,  {Marchetti} L.,  {Cooray} A.,
  {Riechers} D.~A.,    {Amvrosiadis} A.,  2018, \mnras, 476, 4383

\bibitem[\protect\citeauthoryear{{Furlanetto}, {Oh} \& {Briggs}}{{Furlanetto}
  et~al.}{2006}]{2006PhR...433..181F}
{Furlanetto} S.~R.,  {Oh} S.~P.,    {Briggs} F.~H.,  2006, \physrep, 433, 181

\bibitem[\protect\citeauthoryear{{Ghosh}, {Bharadwaj}, {Ali} \&
  {Chengalur}}{{Ghosh} et~al.}{2011}]{2011MNRAS.418.2584G}
{Ghosh} A.,  {Bharadwaj} S.,  {Ali} S.~S.,    {Chengalur} J.~N.,  2011, \mnras,
  418, 2584

\bibitem[\protect\citeauthoryear{{Ghosh}}{{Ghosh}}{2020}]{2020MNRAS.495.2813G}
{Ghosh} A. e.~a.,  2020, \mnras, 495, 2813

\bibitem[\protect\citeauthoryear{{Guha Sarkar}, {Bharadwaj}, {Choudhury} \&
  {Datta}}{{Guha Sarkar} et~al.}{2011}]{2011MNRAS.410.1130G}
{Guha Sarkar} T.,  {Bharadwaj} S.,  {Choudhury} T.~R.,    {Datta} K.~K.,  2011,
  \mnras, 410, 1130

\bibitem[\protect\citeauthoryear{{Gunn} \& {Peterson}}{{Gunn} \&
  {Peterson}}{1965}]{1965ApJ...142.1633G}
{Gunn} J.~E.,  {Peterson} B.~A.,  1965, \apj, 142, 1633

\bibitem[\protect\citeauthoryear{{Gupta}, {Ajithkumar}, {Kale}, {Nayak},
  {Sabhapathy}, {Sureshkumar}, {Swami}, {Chengalur}, {Ghosh},
  {Ishwara-Chandra}, {Joshi}, {Kanekar}, {Lal} \& {Roy}}{{Gupta}
  et~al.}{2017}]{2017CSci..113..707G}
{Gupta} Y.,  {Ajithkumar} B.,  {Kale} H.~S.,  {Nayak} S.,  {Sabhapathy} S.,
  {Sureshkumar} S.,  {Swami} R.~V.,  {Chengalur} J.~N.,  {Ghosh} S.~K.,
  {Ishwara-Chandra} C.~H.,  {Joshi} B.~C.,  {Kanekar} N.,  {Lal} D.~V.,
  {Roy} S.,  2017, Current Science, 113, 707

\bibitem[\protect\citeauthoryear{{Jauzac}, {Richard}, {Jullo}, {Cl{\'e}ment},
  {Limousin}, {Kneib}, {Ebeling}, {Natarajan}, {Rodney}, {Atek}, {Massey},
  {Eckert}, {Egami} \& {Rexroth}}{{Jauzac} et~al.}{2015}]{2015MNRAS.452.1437J}
{Jauzac} M.,  {Richard} J.,  {Jullo} E.,  {Cl{\'e}ment} B.,  {Limousin} M.,
  {Kneib} J.~P.,  {Ebeling} H.,  {Natarajan} P.,  {Rodney} S.,  {Atek} H.,
  {Massey} R.,  {Eckert} D.,  {Egami} E.,    {Rexroth} M.,  2015, \mnras, 452,
  1437

\bibitem[\protect\citeauthoryear{{Jeli{\'c}}, {Zaroubi}, {Labropoulos},
  {Thomas}, {Bernardi}, {Brentjens}, {de Bruyn}, {Ciardi}, {Harker},
  {Koopmans}, {Pandey}, {Schaye} \& {Yatawatta}}{{Jeli{\'c}}
  et~al.}{2008}]{2008MNRAS.389.1319J}
{Jeli{\'c}} V.,  {Zaroubi} S.,  {Labropoulos} P.,  {Thomas} R.~M.,  {Bernardi}
  G.,  {Brentjens} M.~A.,  {de Bruyn} A.~G.,  {Ciardi} B.,  {Harker} G.,
  {Koopmans} L.~V.~E.,  {Pandey} V.~N.,  {Schaye} J.,    {Yatawatta} S.,  2008,
  \mnras, 389, 1319

\bibitem[\protect\citeauthoryear{{Johnson}, {Sharon}, {Bayliss}, {Gladders},
  {Coe} \& {Ebeling}}{{Johnson} et~al.}{2014}]{2014ApJ...797...48J}
{Johnson} T.~L.,  {Sharon} K.,  {Bayliss} M.~B.,  {Gladders} M.~D.,  {Coe} D.,
    {Ebeling} H.,  2014, \apj, 797, 48

\bibitem[\protect\citeauthoryear{{Jullo}, {Kneib}, {Limousin},
  {El{\'\i}asd{\'o}ttir}, {Marshall} \& {Verdugo}}{{Jullo}
  et~al.}{2007}]{2007NJPh....9..447J}
{Jullo} E.,  {Kneib} J.~P.,  {Limousin} M.,  {El{\'\i}asd{\'o}ttir} {\'A}.,
  {Marshall} P.~J.,    {Verdugo} T.,  2007, New Journal of Physics, 9, 447

\bibitem[\protect\citeauthoryear{{Kaiser}}{{Kaiser}}{1987}]{1987MNRAS.227....1K}
{Kaiser} N.,  1987, \mnras, 227, 1

\bibitem[\protect\citeauthoryear{{Kassiola} \& {Kovner}}{{Kassiola} \&
  {Kovner}}{1993}]{1993ApJ...417..450K}
{Kassiola} A.,  {Kovner} I.,  1993, \apj, 417, 450

\bibitem[\protect\citeauthoryear{{Kauffmann}, {White} \&
  {Guiderdoni}}{{Kauffmann} et~al.}{1993}]{1993MNRAS.264..201K}
{Kauffmann} G.,  {White} S.~D.~M.,    {Guiderdoni} B.,  1993, \mnras, 264, 201

\bibitem[\protect\citeauthoryear{{Kulkarni}, {Keating}, {Haehnelt}, {Bosman},
  {Puchwein}, {Chardin} \& {Aubert}}{{Kulkarni}
  et~al.}{2019}]{2019MNRAS.485L..24K}
{Kulkarni} G.,  {Keating} L.~C.,  {Haehnelt} M.~G.,  {Bosman} S. E.~I.,
  {Puchwein} E.,  {Chardin} J.,    {Aubert} D.,  2019, \mnras, 485, L24

\bibitem[\protect\citeauthoryear{{Kumar}, {Dutta} \& {Roy}}{{Kumar}
  et~al.}{2020}]{2020MNRAS.495.3683K}
{Kumar} J.,  {Dutta} P.,    {Roy} N.,  2020, \mnras, 495, 3683

\bibitem[\protect\citeauthoryear{Labate, Braun, Dewdney, Waterson \&
  Wagg}{Labate et~al.}{2017}]{8105424}
Labate M.~G.,  Braun R.,  Dewdney P.,  Waterson M.,    Wagg J.,  2017, in 2017
  XXXIInd General Assembly and Scientific Symposium of the International Union
  of Radio Science (URSI GASS) Ska1-low: Design and scientific objectives.
pp~1--4

\bibitem[\protect\citeauthoryear{{Lanzetta}}{{Lanzetta}}{2000}]{2000eaa..bookE2141L}
{Lanzetta} K.,  2000, {Lyman Alpha Absorption: The Damped Systems}.
p.~2141

\bibitem[\protect\citeauthoryear{{Li} \& {Ostriker}}{{Li} \&
  {Ostriker}}{2002}]{2002ApJ...566..652L}
{Li} L.-X.,  {Ostriker} J.~P.,  2002, \apj, 566, 652

\bibitem[\protect\citeauthoryear{Loutsenko}{Loutsenko}{2018}]{10.1093/ptep/pty119}
Loutsenko I.,  2018, Progress of Theoretical and Experimental Physics, 2018

\bibitem[\protect\citeauthoryear{{Madau}, {Meiksin} \& {Rees}}{{Madau}
  et~al.}{1997}]{1997ApJ...475..429M}
{Madau} P.,  {Meiksin} A.,    {Rees} M.~J.,  1997, \apj, 475, 429

\bibitem[\protect\citeauthoryear{{McQuinn}}{{McQuinn}}{2016}]{2016ARA&A..54..313M}
{McQuinn} M.,  2016, \araa, 54, 313

\bibitem[\protect\citeauthoryear{{Meurer}, {Zheng} \& {de Blok}}{{Meurer}
  et~al.}{2013}]{2013MNRAS.429.2537M}
{Meurer} G.~R.,  {Zheng} Z.,    {de Blok} W.~J.~G.,  2013, \mnras, 429, 2537

\bibitem[\protect\citeauthoryear{{Mitra}, {Choudhury} \& {Ferrara}}{{Mitra}
  et~al.}{2015}]{2015MNRAS.454L..76M}
{Mitra} S.,  {Choudhury} T.~R.,    {Ferrara} A.,  2015, \mnras, 454, L76

\bibitem[\protect\citeauthoryear{{Mo}, {van den Bosch} \& {White}}{{Mo}
  et~al.}{2010}]{2010gfe..book.....M}
{Mo} H.,  {van den Bosch} F.~C.,    {White} S.,  2010, {Galaxy Formation and
  Evolution}

\bibitem[\protect\citeauthoryear{{Modi}, {Castorina}, {Feng} \& {White}}{{Modi}
  et~al.}{2019}]{2019JCAP...09..024M}
{Modi} C.,  {Castorina} E.,  {Feng} Y.,    {White} M.,  2019, \jcap, 2019, 024

\bibitem[\protect\citeauthoryear{{Padmanabhan}}{{Padmanabhan}}{1996}]{1996IAUS..173...55P}
{Padmanabhan} T.,  1996, in {Kochanek} C.~S.,  {Hewitt} J.~N.,  eds,
  Astrophysical Applications of Gravitational Lensing Vol.~173, {Structure
  Formation: Models; Dynamics and Status}.
p.~55

\bibitem[\protect\citeauthoryear{{Peacock} \& {Dodds}}{{Peacock} \&
  {Dodds}}{1994}]{1994MNRAS.267.1020P}
{Peacock} J.~A.,  {Dodds} S.~J.,  1994, \mnras, 267, 1020

\bibitem[\protect\citeauthoryear{{Peebles}}{{Peebles}}{1980}]{1980lssu.book.....P}
{Peebles} P.~J.~E.,  1980, {The large-scale structure of the universe}

\bibitem[\protect\citeauthoryear{{Pen}, {Chang}, {Hirata}, {Peterson}, {Roy},
  {Gupta}, {Odegova} \& {Sigurdson}}{{Pen} et~al.}{2009}]{2009MNRAS.399..181P}
{Pen} U.-L.,  {Chang} T.-C.,  {Hirata} C.~M.,  {Peterson} J.~B.,  {Roy} J.,
  {Gupta} Y.,  {Odegova} J.,    {Sigurdson} K.,  2009, \mnras, 399, 181

\bibitem[\protect\citeauthoryear{{Pourtsidou} \& {Metcalf}}{{Pourtsidou} \&
  {Metcalf}}{2015}]{2015MNRAS.448.2368P}
{Pourtsidou} A.,  {Metcalf} R.~B.,  2015, \mnras, 448, 2368

\bibitem[\protect\citeauthoryear{{Press} \& {Schechter}}{{Press} \&
  {Schechter}}{1974}]{1974ApJ...187..425P}
{Press} W.~H.,  {Schechter} P.,  1974, \apj, 187, 425

\bibitem[\protect\citeauthoryear{{Pritchard} \& {Loeb}}{{Pritchard} \&
  {Loeb}}{2012}]{2012RPPh...75h6901P}
{Pritchard} J.~R.,  {Loeb} A.,  2012, Reports on Progress in Physics, 75,
  086901

\bibitem[\protect\citeauthoryear{{Richard}, {Jauzac}, {Limousin}, {Jullo},
  {Cl{\'e}ment}, {Ebeling}, {Kneib}, {Atek}, {Natarajan}, {Egami}, {Livermore}
  \& {Bower}}{{Richard} et~al.}{2014}]{2014MNRAS.444..268R}
{Richard} J.,  {Jauzac} M.,  {Limousin} M.,  {Jullo} E.,  {Cl{\'e}ment} B.,
  {Ebeling} H.,  {Kneib} J.-P.,  {Atek} H.,  {Natarajan} P.,  {Egami} E.,
  {Livermore} R.,    {Bower} R.,  2014, \mnras, 444, 268

\bibitem[\protect\citeauthoryear{{Richard}, {Kneib}, {Jullo}, {Covone},
  {Limousin}, {Ellis}, {Stark}, {Bundy}, {Czoske}, {Ebeling} \&
  {Soucail}}{{Richard} et~al.}{2007}]{2007ApJ...662..781R}
{Richard} J.,  {Kneib} J.-P.,  {Jullo} E.,  {Covone} G.,  {Limousin} M.,
  {Ellis} R.,  {Stark} D.,  {Bundy} K.,  {Czoske} O.,  {Ebeling} H.,
  {Soucail} G.,  2007, \apj, 662, 781

\bibitem[\protect\citeauthoryear{{Richard}, {Kneib}, {Limousin}, {Edge} \&
  {Jullo}}{{Richard} et~al.}{2010}]{2010MNRAS.402L..44R}
{Richard} J.,  {Kneib} J.~P.,  {Limousin} M.,  {Edge} A.,    {Jullo} E.,  2010,
  \mnras, 402, L44

\bibitem[\protect\citeauthoryear{{Richard}, {Smith}, {Kneib}, {Ellis},
  {Sanderson}, {Pei}, {Targett}, {Sand}, {Swinbank}, {Dannerbauer}, {Mazzotta},
  {Limousin}, {Egami}, {Jullo}, {Hamilton-Morris} \& {Moran}}{{Richard}
  et~al.}{2010}]{2010MNRAS.404..325R}
{Richard} J.,  {Smith} G.~P.,  {Kneib} J.-P.,  {Ellis} R.~S.,  {Sanderson}
  A.~J.~R.,  {Pei} L.,  {Targett} T.~A.,  {Sand} D.~J.,  {Swinbank} A.~M.,
  {Dannerbauer} H.,  {Mazzotta} P.,  {Limousin} M.,  {Egami} E.,  {Jullo} E.,
  {Hamilton-Morris} V.,    {Moran} S.~M.,  2010, \mnras, 404, 325

\bibitem[\protect\citeauthoryear{{Romeo}, {Metcalf} \& {Pourtsidou}}{{Romeo}
  et~al.}{2018}]{2018MNRAS.474.1787R}
{Romeo} A.,  {Metcalf} R.~B.,    {Pourtsidou} A.,  2018, \mnras, 474, 1787

\bibitem[\protect\citeauthoryear{{Saini}, {Bharadwaj} \& {Sethi}}{{Saini}
  et~al.}{2001}]{2001ApJ...557..421S}
{Saini} T.~D.,  {Bharadwaj} S.,    {Sethi} S.~K.,  2001, \apj, 557, 421

\bibitem[\protect\citeauthoryear{{Sarkar} \& {Bharadwaj}}{{Sarkar} \&
  {Bharadwaj}}{2018}]{2018MNRAS.476...96S}
{Sarkar} D.,  {Bharadwaj} S.,  2018, \mnras, 476, 96

\bibitem[\protect\citeauthoryear{{Sarkar}, {Bharadwaj} \&
  {Anathpindika}}{{Sarkar} et~al.}{2016}]{2016MNRAS.460.4310S}
{Sarkar} D.,  {Bharadwaj} S.,    {Anathpindika} S.,  2016, \mnras, 460, 4310

\bibitem[\protect\citeauthoryear{{Sharma}, {Richard}, {Yuan}, {Gupta},
  {Kewley}, {Patr{\'\i}cio}, {Leethochawalit} \& {Jones}}{{Sharma}
  et~al.}{2018}]{2018MNRAS.481.1427S}
{Sharma} S.,  {Richard} J.,  {Yuan} T.,  {Gupta} A.,  {Kewley} L.,
  {Patr{\'\i}cio} V.,  {Leethochawalit} N.,    {Jones} T.~A.,  2018, \mnras,
  481, 1427

\bibitem[\protect\citeauthoryear{{Sharon}, {Bayliss}, {Dahle}, {Dunham},
  {Florian}, {Gladders}, {Johnson}, {Mahler}, {Paterno-Mahler}, {Rigby},
  {Whitaker}, {Akhshik}, {Koester}, {Murray}, {Remolina Gonz{\'a}lez} \&
  {Wuyts}}{{Sharon} et~al.}{2020}]{2020ApJS..247...12S}
{Sharon} K.,  {Bayliss} M.~B.,  {Dahle} H.,  {Dunham} S.~J.,  {Florian} M.~K.,
  {Gladders} M.~D.,  {Johnson} T.~L.,  {Mahler} G.,  {Paterno-Mahler} R.,
  {Rigby} J.~R.,  {Whitaker} K.~E.,  {Akhshik} M.,  {Koester} B.~P.,  {Murray}
  K.,  {Remolina Gonz{\'a}lez} J.~D.,    {Wuyts} E.,  2020, \apjs, 247, 12

\bibitem[\protect\citeauthoryear{{Smith}, {Kneib}, {Smail}, {Mazzotta},
  {Ebeling} \& {Czoske}}{{Smith} et~al.}{2005}]{2005MNRAS.359..417S}
{Smith} G.~P.,  {Kneib} J.-P.,  {Smail} I.,  {Mazzotta} P.,  {Ebeling} H.,
  {Czoske} O.,  2005, \mnras, 359, 417

\bibitem[\protect\citeauthoryear{Swarup}{Swarup}{1991}]{swarup...1991}
Swarup G.,  1991, International Astronomical Union Colloquium, 131, 376–380

\bibitem[\protect\citeauthoryear{{Swarup}, {Sarma}, {Joshi}, {Kapahi}, {Bagri},
  {Damle}, {Ananthakrishnan}, {Balasubramanian}, {Bhave} \& {Sinha}}{{Swarup}
  et~al.}{1971}]{1971NPhS..230..185S}
{Swarup} G.,  {Sarma} N.~V.~G.,  {Joshi} M.~N.,  {Kapahi} V.~K.,  {Bagri}
  D.~S.,  {Damle} S.~H.,  {Ananthakrishnan} S.,  {Balasubramanian} V.,  {Bhave}
  S.~S.,    {Sinha} R.~P.,  1971, Nature Physical Science, 230, 185

\bibitem[\protect\citeauthoryear{{Tegmark}, {Silk}, {Rees}, {Blanchard}, {Abel}
  \& {Palla}}{{Tegmark} et~al.}{1997}]{1997ApJ...474....1T}
{Tegmark} M.,  {Silk} J.,  {Rees} M.~J.,  {Blanchard} A.,  {Abel} T.,
  {Palla} F.,  1997, \apj, 474, 1

\bibitem[\protect\citeauthoryear{{Trott}}{{Trott}}{2016}]{2016ApJ...818..139T}
{Trott} C.~M. e.~a.,  2016, \apj, 818, 139

\bibitem[\protect\citeauthoryear{{Villaescusa-Navarro}, {Viel}, {Datta} \&
  {Choudhury}}{{Villaescusa-Navarro} et~al.}{2014}]{2014JCAP...09..050V}
{Villaescusa-Navarro} F.,  {Viel} M.,  {Datta} K.~K.,    {Choudhury} T.~R.,
  2014, \jcap, 2014, 050

\bibitem[\protect\citeauthoryear{{Weinberger}, {Kulkarni} \&
  {Haehnelt}}{{Weinberger} et~al.}{2020}]{2020MNRAS.494..703W}
{Weinberger} L.~H.,  {Kulkarni} G.,    {Haehnelt} M.~G.,  2020, \mnras, 494,
  703

\bibitem[\protect\citeauthoryear{White \& Rees}{White \&
  Rees}{1978}]{10.1093/mnras/183.3.341}
White S. D.~M.,  Rees M.~J.,  1978, Monthly Notices of the Royal Astronomical
  Society, 183, 341

\bibitem[\protect\citeauthoryear{{Wootten} \& {Thompson}}{{Wootten} \&
  {Thompson}}{2009}]{2009IEEEP..97.1463W}
{Wootten} A.,  {Thompson} A.~R.,  2009, IEEE Proceedings, 97, 1463

\bibitem[\protect\citeauthoryear{{Yoshiura}, {Line}, {Kubota}, {Hasegawa} \&
  {Takahashi}}{{Yoshiura} et~al.}{2018}]{2018MNRAS.479.2767Y}
{Yoshiura} S.,  {Line} J.~L.~B.,  {Kubota} K.,  {Hasegawa} K.,    {Takahashi}
  K.,  2018, \mnras, 479, 2767

\bibitem[\protect\citeauthoryear{{Zaroubi}}{{Zaroubi}}{2013}]{2013ASSL..396...45Z}
{Zaroubi} S.,  2013, {The Epoch of Reionization}.
p.~45

\bibitem[\protect\citeauthoryear{{Zhang}, {Bunn}, {Karakci}, {Korotkov},
  {Sutter}, {Timbie}, {Tucker} \& {Wandelt}}{{Zhang}
  et~al.}{2016}]{2016ApJS..222....3Z}
{Zhang} L.,  {Bunn} E.~F.,  {Karakci} A.,  {Korotkov} A.,  {Sutter} P.~M.,
  {Timbie} P.~T.,  {Tucker} G.~S.,    {Wandelt} B.~D.,  2016, \apjs, 222, 3

\bibitem[\protect\citeauthoryear{{Zhou}, {Wu}, {Zhou} \& {Ma}}{{Zhou}
  et~al.}{2018}]{2018PASP..130i4101Z}
{Zhou} Z.,  {Wu} H.,  {Zhou} X.,    {Ma} J.,  2018, \pasp, 130, 094101

\end{thebibliography}

\end{document}